%% file: main_arXiv.tex
\documentclass{article}

\input{./FileSettings/SetupMain}
\usepackage{tikzit}
\input{paper_styles.tikzstyles}
\usetikzlibrary{calc}
\usetikzlibrary{arrows.meta}
\usetikzlibrary{calc}

\begin{document}
\input{./Contents/TitlePage}
\input{./Contents/MainContents}

\newpage
\bibliography{references_arXiv}

\end{document}

%% file: FileSettings/SetupMain.tex
\input{FileSettings/MainPackages}
\input{FileSettings/Definitions}
\input{FileSettings/BibSettings}

%% file: FileSettings/MainPackages.tex
\usepackage{FileSettings/arxiv}
\usepackage[utf8]{inputenc} 
\usepackage[T1]{fontenc}    
\usepackage{url}  
\usepackage{soul}
\usepackage{booktabs}       
\usepackage{amsfonts}       
\usepackage{amsmath}
\usepackage{nicefrac}       
\usepackage{multirow}
\usepackage{pifont}
\usepackage{caption} 
\usepackage{subcaption}
\usepackage{microtype}      
\usepackage{lipsum}		
\usepackage{amsmath}
\usepackage{graphicx}
\usepackage{physics}
\usepackage{listings}
\usepackage[table]{xcolor}
\usepackage[pdftex,colorlinks=true, allcolors=blue, linkcolor=blue]{hyperref}
\usepackage{xcolor}
\usepackage[most]{tcolorbox} 
\usepackage[framemethod=tikz]{mdframed}
\usetikzlibrary{positioning}
\usepackage[labelfont=bf]{caption}
\usepackage{hyperref}
\usepackage{tabularx}

%% file: FileSettings/Definitions.tex
\newcommand{\headeright}{A preprint}
\newcommand{\undertitle}{A preprint}

\definecolor{codegreen}{rgb}{0,0.6,0}
\definecolor{codegray}{rgb}{0.5,0.5,0.5}
\definecolor{codepurple}{rgb}{0.58,0,0.82}
\definecolor{backcolour}{rgb}{0.95,0.95,0.92}

\lstdefinestyle{mystyle}{
  commentstyle=\color{codegreen},
  keywordstyle=\color{magenta},
  numberstyle=\tiny\color{codegray},
  stringstyle=\color{codepurple},
  basicstyle=\ttfamily\footnotesize,
  breakatwhitespace=false,         
  breaklines=true,                 
  captionpos=b,                    
  keepspaces=true,                 
  numbers=left,                    
  numbersep=10pt,                  
  showspaces=false,                
  showstringspaces=false,
  showtabs=false,                  
  tabsize=2,
  xleftmargin=5pt,
  aboveskip=5pt,
  belowskip=0pt
}

\definecolor{light-gray}{gray}{0.95}

\surroundwithmdframed[
  hidealllines=true,
  backgroundcolor=light-gray,
  innerleftmargin=0pt,
  skipabove=10pt,
  skipbelow=20pt]{lstlisting}

\definecolor{HeaderGray}{gray}{0.85}	
\definecolor{Gray}{gray}{0.9}
\definecolor{LightCyan}{rgb}{0.78,0.97,0.97}

\DeclareMathOperator{\sgn}{sgn}
\DeclareMathOperator{\rand}{rand_U}

\newcommand{\subfigref}[2]{\hyperref[{#1}]{\ref*{#1}#2}}

%% file: paper_styles.tikzstyles
\tikzstyle{new style 0}=[fill={rgb,255: red,160; green,255; blue,108}, draw=black, shape=circle]

\tikzstyle{boxes}=[-, fill={rgb,255: red,191; green,191; blue,191}, rounded corners=0.2cm, draw={rgb,255: red,191; green,191; blue,191}]

%% file: Contents/TitlePage.tex
\title{A virtually connected probabilistic computer as a solver for higher-order, densely connected, or reconfigurable combinatorial optimisation problems}

\author{\href{https://orcid.org/0000-0003-1795-871X}{
        \includegraphics[scale=0.06]{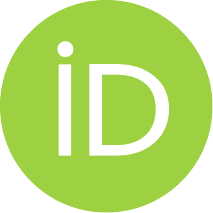}\hspace{1mm}Amy J. Searle}\textsuperscript{1,*} \and
        \href{https://orcid.org/0009-0007-3004-6367}{
        \includegraphics[scale=0.06]{Figures/Misc/orcid.pdf}\hspace{1mm}Harry Youel}\textsuperscript{1,2,3} \and 
        \href{https://orcid.org/0009-0007-9273-4480}{
        \includegraphics[scale=0.06]{Figures/Misc/orcid.pdf}\hspace{1mm}Fredrik Hasselgren}\textsuperscript{1,4} \and \href{https://orcid.org/0000-0002-2056-6437}{\includegraphics[scale=0.06]{Figures/Misc/orcid.pdf}\hspace{1mm}Annika Möslein}\textsuperscript{1} \and 
        \href{https://orcid.org/0000-0002-9871-4962}{\includegraphics[scale=0.06]{Figures/Misc/orcid.pdf}\hspace{1mm}Ramy Aboushelbaya}\textsuperscript{1} \and \href{https://orcid.org/0000-0002-7192-0256}{\includegraphics[scale=0.06]{Figures/Misc/orcid.pdf}\hspace{1mm}Marko von der Leyen}\textsuperscript{1,**}}

\date{}

\maketitle

\begin{center}
    \textsuperscript{1}Quantum Dice Limited, 264 Banbury Road,\\ Oxford, OX2 7DY, United Kingdom \\
    
    \textsuperscript{2}London Centre for Nanotechnology, University College London,\\ London, WC1H 0AH, United Kingdom \\

    \textsuperscript{3}Department of Physics and Astronomy, University College London,\\ London, WC1E 6BT, United Kingdom\\

    \textsuperscript{4}Mathematical Institute, University of Oxford,\\ Oxford, OX2 6GG, United Kingdom\\

    \vspace{0.25cm}\textsuperscript{*}\href{mailto:amyjadesearle@gmail.com}{\texttt{amyjadesearle@gmail.com}},
    \textsuperscript{**}\href{mailto:marko.leyen@quantum-dice.com}{\texttt{marko.leyen@quantum-dice.com}}
    
\end{center}

\vspace{1cm}

\input{Contents/Abstract}
\keywords{Probabilistic computing, photonics, combinatorial optimisation, set cover, travelling salesperson problem, spin-glass, Max-Cut, Ising model, QUBO}

%% file: Contents/Abstract.tex
\begin{abstract}
Recently, there has been growing interest in unconventional computing as an approach for solving NP-hard problems, by developing dedicated hardware to find solutions more efficiently than conventional CPUs. In many of these approaches, however, certain problem geometries must be transformed into forms that are more amenable to the available hardware topology through techniques such as embedding, sparsification, and quadratisation, leading to a deterioration in solution quality. A probabilistic computing architecture based on high speed photonic quantum random number generators was recently proposed which utilises virtual hardware connections (Aboushelbaya \textit{et al.}, $2025$), circumventing the necessity for such procedures. Here, we discuss the applicability of virtually connected hardware for running heuristic solving methods to solve a selection of problems, which due to their geometry, would suffer from topological hardware restrictions. We also employ greedy graph colouring algorithms for hardware parallelisation, allowing favourable scaling for desirable solution qualities. To emphasise the difficulty in solving these problems on physically connected hardware, we demonstrate the increase in problem size that would occur with quadratisation or sparsification. Using simulations to emulate hardware, we predict that a photonic probabilistic computer would outperform the time to solution recently reported for digital annealing units, on the ground state approximation of Erdős–Rényi graph spin-glasses, by orders of magnitude.
\end{abstract}

%% file: Contents/MainContents.tex
\input{Contents/Introduction}
\input{Contents/Overview}
\input{Contents/NPProblems}
\input{Contents/Conclusion}
\input{Contents/Contributions}
\input{Contents/Acknowledgments}
\input{Contents/Appendices}

%% file: Contents/Introduction.tex
\section{Introduction}
The Ising model, and the equivalent class of quadratic unconstrained binary optimisation (QUBO) problems, act as important test-beds for solving many combinatorial optimisation problems; appropriate choices for the quadratic and linear terms enable the solution to a specified optimisation problem to be encoded in the ground state of a model~\cite{Lucas_2014}. Typically, Monte-Carlo sampling is employed with a global temperature variable to maximise the probability of reaching the ground state, with simulated annealing (SA) and parallel tempering (PT) being common examples of such algorithms.

Besides the simplicity of using a single energy function formulation to capture a number of diverse optimisation problems, the Ising formulation is attractive because of the possibility of using dedicated hardware to implement such algorithms. The purpose of this is to achieve a large reduction in the time to solution (TTS) for a given problem~\cite{Patel2020IsingMO, Leleu_2021}, often in cases where exact answers are not required, and there is a preference for approximate solutions in reduced runtimes. This form of computing, coined ‘Nature-inspired’ and first conceptualised in 1982 by Feynman~\cite{Feynman_1982}, has motivated the development of hardware including quantum annealers~\cite{Hayato_2017}, digital ASIC annealers~\cite{Aramon_2019}, oscillator networks~\cite{Moy_2022}, and probabilistic Ising machines. The latter has been implemented in various forms, including using digital CMOS technology~\cite{Finocchio_2021} and magnetic tunnel junctions (MTJs)~\cite{Grollier_2020}, which employ interconnected binary stochastic neurons to drive computation~\cite{Camsari_2017}.

Formulating the Ising model, however, involves specifying a quadratic energy function, consequently only allowing for dyadic or pairwise interactions. This practically restricts us from implementing problems which require higher-order energy functions without the introduction of auxiliary variables which significantly increases the search space of the problem, making energy landscapes less favourable for heuristic algorithms to explore~\cite{Valiante_2021}. Moreover, practical hardware implementations often impose additional constraints on the interaction density, particularly in quantum annealers, but also in CMOS or MTJ-based probabilistic computing, since realising dense, all-to-all connectivity at scale presents a large engineering challenge. One avenue has proposed reconfigurable architectures, but their approach was only valid in the limit of sparse interactions~\cite{Nikhar_2025}. Quantum annealers aim to circumvent the issue by embedding the problem of interest into the architecture of the hardware, but since such embeddings are expensive to compute, usually a universal but inefficient embedding is used~\cite{Okada_2019}. Other platforms offer an advantage in this regard, such as digital annealers that support all-to-all connectivity~\cite{Aramon_2019}, or recently developed hardware in probabilistic computing, which leverages photonic quantum random number generators as an entropy source for probabilistic bits and implements their connections virtually~\cite{Ramy_2025}. These virtual connections are of particular interest, as we envision that this could simplify the hardware realisation of three types of problem: dense quadratic problems, reconfigurable problems (where the geometry changes dynamically), and higher-order problems. Reconfigurable problems present a challenge because multiple embeddings must occur sequentially, increasing the complexity of the algorithm. Following this observation, we study three problems, each of which constitutes a difficult geometry for distinct reasons. First, the set cover problem presents higher-order interactions; it belongs to a general class of cover problems with applications ranging from crew scheduling to search algorithms~\cite{Caprara_2000, Beasley_1987}. Second, the travelling salesperson problem (TSP) requires a reconfigurable geometry when solved with $k$-means clustering (KMC) preprocessing. TSP is a widely studied combinatorial optimisation task with applications in logistics, route planning~\cite{Gohil_2022, Bock_2025}, genomics, and astronomy~\cite{Osaba_2020, Handley_2023}. Finally, spin-glasses on Erdős–Rényi graphs necessitate dense interactions; this problem originates in condensed matter physics but has implications for tasks ranging from network flow to job scheduling~\cite{Kochenberger_2014}.

We begin this study by demonstrating the performance of a simulated virtually connected probabilistic computer (VCPC) across these three problem classes, evaluating the best attainable solution qualities and the iteration count required to reach specific target qualities by employing both SA and linear PT algorithms. Furthermore, we quantify the impact of hardware constraints by demonstrating how sparsification and quadratisation lead to a significant increase in problem size. For the higher-order set cover problem, we show that this leads to a clear deterioration in solution quality for a fixed number of iterations. Regarding the spin-glass problem, we solve the NP-complete weighted Max-Cut problem on Erdős–Rényi graphs with edge weights randomly chosen from $\{-1, 1\}$, making the results directly relevant to a broad class of computational tasks, that can be expressed through NP-complete reductions~\cite{karp_reducibility_1972}. We compare our estimated VCPC performance to findings reported in Ref.~\cite{Aramon_2019}, arguing that the predicted TTS on the hardware introduced in Ref.~\cite{Ramy_2025} would achieve orders of magnitude improvements over those reported for digital annealing.

The remainder of this paper is structured as follows. We provide an overview of probabilistic computing in Sec.~\ref{sec:overview} and present our main findings in Sec.~\ref{sec:problems}. In particular, we detail the ability of the VCPC to solve the set cover, TSP, and spin-glass problems, highlighting the advantage of virtual connectivity for problems presenting complex geometries that would otherwise necessitate prohibitive overheads in physically constrained hardware.

%% file: Contents/Overview.tex
\section{An overview of probabilistic computing}
\label{sec:overview}
A probabilistic computer consists of stochastic binary units, termed probabilistic bits (p-bits), which are interconnected to form a network. We can associate an energy function with a given network by tuning the network's interactions to equate the p-bits' states with energy contributions. In the case of a quadratic energy function, a probabilistic computer represents an Ising model. The stochasticity of a p-bit is controlled by it's input, $I$, and the corresponding output $s$, fluctuates between $-1$ and $1$ with a probability determined by $I$:
\begin{equation}
    s = \sgn \{\sigma(\beta I) + \rand[-1,+1]\},
\end{equation}
where $\sigma$ is the sigmoid function, $\beta$ is the inverse temperature, $\rand[-1,+1]$ is a uniform random number on the interval $[-1,+1]$. Bipolar outputs can be converted to binary outputs via the mapping  $\frac{s}{2} +\frac{1}{2}$~\cite{Camsari_2017}. The input of a given p-bit is determined by the outputs of it's neighbours in the network. When such interactions are pairwise, the input of the $i$th p-bit is computed as 
\begin{equation}
    I_i = \sum_j J_{ij} s_j + h_i,
    \label{eqn:bias-calculation}
\end{equation}
where $J_{ij}$ is the coupling matrix describing, the interaction between p-bits $i$ and $j$, and $h_i$ is the constant bias contribution. We also refer to an input $I_{i}$ as the $i^{\text{th}}$ p-bit's update drive when solving a specific problem, which we detail in App.~\ref{app:bias-eqns}. In this work, we will also consider higher-order networks, which for three-way interactions can be represented as:
\begin{equation}
    I_i = \sum_{jk} J^3_{ijk} s_{j}s_{k} + \sum_{j} J^2_{ij} s_{j} + h_i,
\end{equation}
and can be generalised to higher-order polynomials. 
The network will, if $J$ is symmetric, after a sufficient number of p-bit updates, be described by the Boltzmann distribution with state probabilities: 
\begin{equation}\label{eqn:boltz-prob}
    p(E(\mathbf{s})) = \frac{e^{- \beta E(\mathbf{s})}}{Z}
\end{equation}
where the energy $E$ of a configuration of p-bit states $\mathbf{s} = (s_1, s_2, ..., s_n)$ is given by
\begin{equation}
    E(\mathbf{s}) = - \sum_{i<j} J_{ij} s_i s_j - \sum_i h_i s_i,
\end{equation}
and $Z$ is the partition function. This energy function can be expressed in QUBO form via the Ising to QUBO mapping that we show explicitly in App.~\ref{app:SG-Mapping}. Again, this can be generalised to higher-order interacting systems -- three-way and beyond. In this case the energy function is known as a higher-order unconstrained binary optimisation problem (HUBO).

A system which natively approximates the Boltzmann distribution is useful for optimisation, because encoding the solution to a problem in the ground state enables the system naturally visit the optimal state with the highest probability, in accordance with the relation in Eqn.~\ref{eqn:boltz-prob}. There are a number of algorithms, to be discussed in the next section, which have been developed in the field of energy-based optimisation to assist the system in finding the global ground state, by avoiding convergence to local minima. 

The update drive in Eqn.~\ref{eqn:bias-calculation} can be determined by calculating the energy difference between the zero and one states of the $i^{\text{th}}$ p-bit, and is expressed as: 
\begin{equation}
    I_i = E(\mathbf{s} \vert s_i = 0) - E(\mathbf{s} \vert s_i = 1).
    \label{eqn:bias-energy-diff}
\end{equation}

While Eqn.~\ref{eqn:bias-calculation} implies that the bias of a p-bit is updated through its direct interaction with neighbouring p-bits in the network, it is also possible to mimic such interactions with virtual connections. By supplementing p-bits with a central computing unit which is able to read the p-bits' states required to perform the computation in Eqn.~\ref{eqn:bias-calculation} or Eqn.~\ref{eqn:bias-energy-diff}, and is also able to bias p-bit $i$, we can avoid directly implementing physical connections between p-bits. An in-depth discussion of a photonic probabilistic computer which utilises virtual connections in this manner was recently introduced in Ref.~\cite{Ramy_2025}, and a brief discussion is provided in App.~\ref{app:hardware_strategies}.

\subsection{Optimisation methods}
There are a number of approaches to increase the solution quality of Boltzmann-like architectures, including SA; PT, which may or may not be adaptive; and simulated quantum annealing. In this work we explore SA and non-adaptive PT.

\textit{Simulated annealing}: A standard optimisation algorithm, where an external temperature variable is introduced, and slowly decreased. At high temperatures, the system samples large subsections of its configuration space, while at low temperatures the system is less likely to move states, and samples smaller subspaces. Conventionally, we define $\beta$ to denote to the inverse temperature of the system. We let $\beta_{\mathrm{start}}$ and $\beta_{\mathrm{end}}$ indicate the start and end points of this interval, and $S$ indicate the number of steps in this interval. These steps are linearly separated within the interval; although one can alternatively use geometric spacing.  At each step, a p-bit group is chosen at random $I$ times, and the update drives of the incumbent p-bits are calculated in parallel. This can be done if p-bits in a group are independent, meaning that the state of one p-bit is not needed to calculate the update drive of another p-bit (see Eqn.~\ref{eqn:bias-calculation}). Usually, finding these groups is itself and NP-hard problem since it is equivalent to a graph or hypergraph colouring problem, but we employ greedy algorithms which still offer dramatic speed-up through this group parallelisation. Since the algorithm is probabilistic, we repeat the run $R$ times. 

\textit{Parallel tempering}: Another optimisation algorithm, which involves introducing $P$ copies of the system at different temperatures, referred to as replicas, and probabilistically exchanging the states between two adjacent replicas periodically. The probability of swapping two replicas' states is given by the Metropolis criteria:
\begin{equation}
    P_{i, i+1} = \min \{1, \exp{(\Delta \beta \Delta E)} \},
\end{equation}
where $\Delta \beta = \beta_{i+1} - \beta_i$ and $\Delta E = E_{i+1} - E_i$ are the differences in inverse temperature and energy between neighbouring replicas, respectively. This ensures that low temperature states do not get stuck in local minima, by propagating other potentially promising states from higher temperatures. The PT algorithm is often favoured in the probabilistic computing literature \cite{Grimaldi_2022, raimondo2025}, although there is not yet a thorough understanding of when PT or SA is favourable. Nevertheless in certain cases PT has been shown to perform better, such as on Ising spin-glasses with Gaussian distributed couplings \cite{Rom_2009}. Here, we employ linear PT, where the replicas' temperatures are linearly interpolated between $(\beta_{\text{start}}, \beta_{\text{end}})$. Similarly to SA, we perform $I$ iterations on each replica, swapping replicas at an interval of $\mathcal{S}$. Again, at each iteration a group of p-bits is selected and their update drives are calculated in parallel. 

We refer to SA and PT on the simulated VCPC as PC-SA and PC-PT respectively, to highlight that we refer to the algorithms described above.

\subsection{Scaling of the heuristic probabilistic algorithm}
We now give a description of our probabilistic algorithm's complexity, and highlight the speed-ups gained by parallelisation. Considering SA, suppose we have $S$ temperature steps with $I$ iterations at each step (let the total number of iterations be $\mathcal{I} = I \times S$). Without any parallelisation, suppose that the number of iterations needed to reach a solution of quality $q$ with probability $p$ is $\mathcal{I}_{q,p}$. Then, the total number of operations is
\begin{equation}\label{eqn:complexity}
    O_{q,p} = \mathcal{I}_{q,p} \cdot \mathcal{O}(p({\mathbf{n}}))
\end{equation}
where the prefactor indicates the number of times an update drive calculation is completed, and $\mathcal{O}(p({\mathbf{n}}))$ indicates that the update drive calculation scales at some polynomial $p({\mathbf{n}})$ that is dependent on the problem parameters ${\mathbf{n}}$. Now suppose we run $R$ repeats, and, further, that we group p-bits into a list of groups $G$ such that any set $g \in G$ contains only p-bits which do not occur in their update drive calculation. Then, 
\begin{equation}\label{eqn:complexity_group}
    O_{q,p} = \frac{\mathcal{I}_{q,p} \cdot \mathcal{O}(p({\mathbf{n}})) \cdot R}{\bar{G}}.
\end{equation}
Assuming we are able to run $R$ repeats in parallel, this simply becomes
\begin{equation}\label{eqn:complexity_group2}
    O_{q,p} = \frac{\mathcal{I}_{q,p} \cdot \mathcal{O}(p({\mathbf{n}}))}{\bar{G}}.
\end{equation}
where $\bar{G}$ is the average size of groups in $G$. If every group is a singleton (only containing one p-bit), we have $\bar{G} = 1$ and no speed-up is obtained. Note, therefore, that the scaling of the group sizes is extremely important for obtaining speed-ups when implementing these algorithms, as well as the form of the polynomial $p({\mathbf{n}})$. We may also express this in terms of the number of groups in the set $G$, since $|G|/N = 1/\bar{G}$. It was recently shown by a subset of the authors, that perfect parallelisation in this sense can result in favourable scaling in cubic and biclique geometries for the spin-glass problem~\cite{Fredrik_2025}.

Note that although parallelisation is important for ensuring the algorithm runs efficiently, such parallelisation procedures are not specific to the hardware and can be done also on CPUs and GPUs. The expected speed-up when implementing this on a VCPC originates from increases in sampling speed, which is discussed in Ref.~\cite{Chowdhury_2025}. Nevertheless, parallelisation is an important consideration for reducing the runtime of probabilistic algorithms.

\subsection{Advantages of a virtually connected probabilistic computer}

As already emphasised, hardware dedicated to heuristic optimisation is often restricted to QUBO formulated energy functions:
\begin{equation}
    E(\mathbf{s}) = -\sum_{\substack{i, j \in V \\ i <j}} J_{i,j} s_is_j - \sum_{i \in V} h_i s_i.
    \label{eqn:qubo-energy}
\end{equation}
Here, we use $\mathbf{s} = (s_1,s_2,...,s_n)$ to emphasise that this is the general case, without reference to probabilistic computing. We may encounter optimisation problems which are represented by higher dimensional energy functions:
\begin{equation}
    E(\mathbf{s}) = -\sum_{\substack{i^1, i^2, ..., i^k \in V\\i^1 < i^2 ... < i^k}} J^{k}_{i^1, i^2, ..., i^k} s_{i^1}s_{i^2}...s_{i^k} - \sum_{\substack{i^1, i^2, ..., i^{k-1} \in V\\i^1 < i^2 <...< i^{k-1}}} J^{k-1}_{i^1, i^2, ..., i^{k-1}} s_{i^1}s_{i^2}...s_{i^{k-1}} - ... - \sum_{i \in V} h_{i} s_{i}.
    \label{eqn:energy-higher-order}
\end{equation}
Here, we refer to $k$ as the dimension of the problem. QUBOs and Ising models are therefore of dimension $2$, and a dimension $k$ problem contains $k$ tensors in the Hamiltonian from dimension $k$ to dimension $1$ (the $h$ vector). For a finite $J^l_{i^1, i^2, ..., i^l}$, the corresponding entities $s_{i^1}$ through to $s_{i^k}$ share a non-zero $l$--way interaction.

\begin{figure}
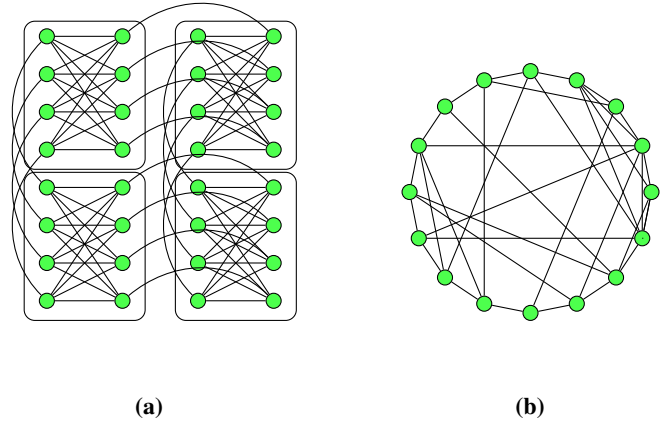

\centering
\begin{subfigure}{.3 \linewidth}
\centering
    \include{Figures/Main/dwave_arc}
    \caption{}
    \label{subfig:dwave-arc}
\end{subfigure}
\begin{subfigure}{.3 \linewidth}
\centering
    \include{Figures/Main/sparse-arc}
    \caption{}
    \label{subfig:sparse-arc}
\end{subfigure}

    \caption{\textbf{(a)} The Chimera architecture of D-Wave Inc., a quantum annealer which is often employed to demonstrate quantum annealing performance. In order to represent higher-order energy functions as a Chimera graph, or other imposed architectures, an expensive quadratisation process must be performed. When converting a densely connected problem, embedding or sparsification processes must also be employed. \textbf{(b)} A random sparse architecture. The process of sparsification transforms a dense graph into a sparse graph, by reducing each vertex's number of allowed connections.}
    \label{fig:hardware-geometries}
\end{figure}

Typically, hardware is restricted to certain interaction geometries; for D-Wave quantum annealers this is a Chimera graph, as shown in Fig.~\ref{subfig:dwave-arc}. More generally, at least in the case that only interactions of order two are implemented, the allowed interaction geometry of a device is represented by a graph $G$, which is defined by an adjacency matrix $A_{ij}$. Entries where $A_{ij}=0$ indicate that the hardware is not able to directly implement an interaction between $s_i$ and $s_j$. As such, anytime that (i) interactions are of order higher than $2$, or (ii) $J_{ij}$ is non-zero but $A_{ij} =0$, requires additional pre-processing to transform the problem into a representation that the hardware can manage. The down-side of these processes is that they introduce additional variables, called auxiliary variables, effectively increasing the resources required to perform the minimisation. The processes that deal with (i) and (ii), respectively, are quadratisation, and embedding or sparsification. While embedding directly finds a map to target hardware, sparsification, which is closely related to embedding, has no target topology in mind and simply aims to reduce the number of neighbours that each vertex has at the cost of increasing the number of vertices. An example of a sparse random graph (random here referring to the fact that edges are randomly placed) is shown in Fig.~\ref{subfig:sparse-arc}.
\begin{enumerate}
    \item \textit{Quadratisation} is the process of turning a higher-order energy function into a quadratic energy function. Note that usually, after this step, an embedding into the hardware graph still needs to be implemented. There is a standard process for turning higher-order terms into quadratic terms, where a term of order $k$ introduces $k-2$ additional terms. As such, if there are $n$ terms of order $k$, $n(k-2)$ auxiliary variables are needed. It is known that quadratisation causes not only an increased in the size of the problem, but also results in a rugged energy landscape~\cite{Dobrynin_2024}.
    \item \textit{Embedding} is the process of mapping the logical variables and pairwise interactions of a problem to the physical qubits and couplers of the hardware interactions graph $G$. Typically the hardware graph is sparse, so finding an embedding, often requiring groups of physical qubits to represent a single logical variable, is computationally expensive and consumes a large number of physical qubits. This overhead drastically limits the effective problem size; for instance a 1000-qubit Chimera processor may only accommodate a problem of $\sim 45$ variables~\cite{Albash}. As the number of required number of physical qubits scales poorly with the problem size, it is rarely feasible to embed large or dense problems directly. Instead, the problem is broken down into smaller subproblems, each of which is solved separately~\cite{Pelofske_2019}.
    \item \textit{Sparsification} is the process of reducing the connectivity of a dense problem's interaction graph to a sparse, energetically degenerate surrogate~\cite{Sajeeb_2025}. Fewer neighbouring interactions per vertex are desirable as in general graphs with lower average connectivity (where connectivity refers to the number of vertices a single vertex is connected to) are easier to implement in hardware and can be coloured using less update groups. The drawback of this approach is that it leads to a significant expansion in the number of variables relative to the original problem formulation.
\end{enumerate}

%% file: Figures/Main/dwave_arc.tex
\begin{tikzpicture}[scale=1.0, every node/.style={circle, draw, fill=green!70,inner sep=2pt}]

  \def\cellsize{2.0}   
  \def\n{2}            

  \foreach \i in {0,...,\numexpr\n-1} {
    \foreach \j in {0,...,\numexpr\n-1} {
      \foreach \k in {0,...,3} {
        \node (L\i\j\k) at (\i*\cellsize, \j*\cellsize+0.5*\k) {};
      }
      \foreach \k in {0,...,3} {
        \node (R\i\j\k) at (\i*\cellsize+1, \j*\cellsize+0.5*\k) {};
      }
      \foreach \k in {0,...,3} {
        \foreach \m in {0,...,3} {
          \draw (L\i\j\k) -- (R\i\j\m);
        }
      }
      \draw[rounded corners] 
        (\i*\cellsize-0.3, \j*\cellsize-0.26) rectangle 
        (\i*\cellsize+1.3, \j*\cellsize+0.3+0.5*2.8);
    }
  }

\draw [out=135,in=225] (L003) edge (L013);
  \draw [out=135,in=225] (L002) edge (L012);
  \draw [out=135,in=225] (L001) edge (L011);
  \draw [out=135,in=225] (L000) edge (L010);
  

    \draw [out=135,in=225] (L103) edge (L113);
  \draw [out=135,in=225] (L102) edge (L112);
  \draw [out=135,in=225] (L101) edge (L111);
  \draw [out=135,in=225] (L100) edge (L110);






    \draw [out=135,in=40] (R113) edge (R013);
    \draw [out=135,in=40] (R112) edge (R012);
    \draw [out=135,in=40] (R111) edge (R011);
    \draw [out=135,in=40] (R110) edge (R010);



    \draw [out=135,in=40] (R100) edge (R000);
    \draw [out=135,in=40] (R101) edge (R001);
    \draw [out=135,in=40] (R102) edge (R002);
    \draw [out=135,in=40] (R103) edge (R003);

\end{tikzpicture}

%% file: Figures/Main/sparse-arc.tex
\begin{tikzpicture}[scale=0.4, 
  vertex/.style={circle,draw,fill=green!70,inner sep=2pt},
  edge/.style={}
]

  \def\N{16}           
  \def\R{4}            
  \def\pExtra{0.15}    
  \pgfmathsetseed{2468} 

  \foreach \i in {1,...,\N}{
    \path coordinate (p\i) at ({\R*cos(360*(\i-1)/\N)}, {\R*sin(360*(\i-1)/\N)});
    \node[vertex] (v\i) at (p\i) {};
  }

  \foreach \i in {1,...,\N}{
    \pgfmathtruncatemacro{\j}{mod(\i,\N)+1} 
    \draw[edge] (v\i) -- (v\j);
  }

  \foreach \i in {1,...,\N}{
    \pgfmathtruncatemacro{\istart}{\i+2} 
    \foreach \j in {\istart,...,\N}{
      \pgfmathsetmacro{\u}{rnd}
      \ifdim \u pt<\pExtra pt
        \draw[edge] (v\i) -- (v\j);
      \fi
    }
  }

\end{tikzpicture}

%% file: Contents/NPProblems.tex
\section{Problems}
\label{sec:problems}
Here we outline three problems which we solve to evaluate the performance of the VCPC algorithm. Each problem is known to be NP-hard and presents different challenges in hardware realisation. 

\subsection{Set cover}
\begin{figure}[h]
    \centering
    \begin{subfigure}{.4\linewidth}\centering
    \includegraphics[width = 0.9 \textwidth]{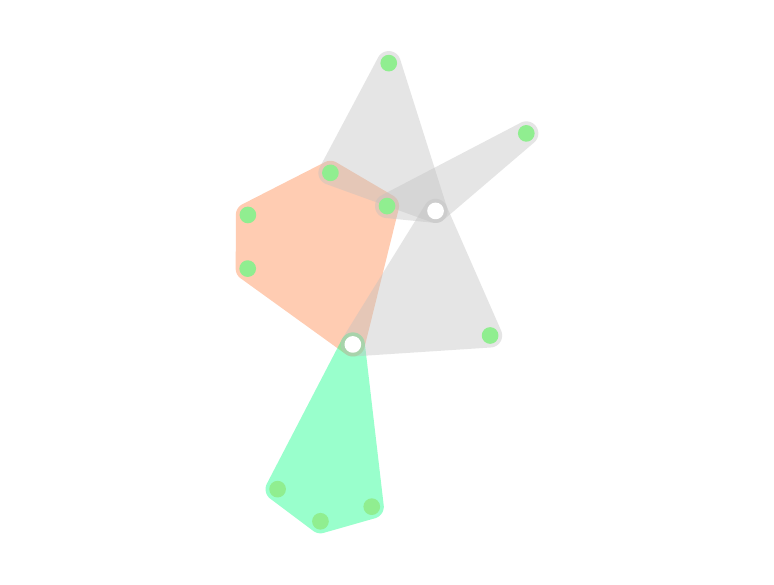}
    \caption{}\label{subfig:hypergraph}
    \end{subfigure}
    \begin{subtable}{.45\linewidth}\centering
    {\begin{tabular}{c|ccccc}
          & $V_1$ & { \fcolorbox{gray}{yellow}{$V_2$}} & $V_3$ & ... &{ \fcolorbox{gray}{yellow}{$V_n$}}\\[2mm]
          \hline 
          
        $0$ & 0 &  { \fcolorbox{gray}{yellow}{1}} & 0 & ... & { \fcolorbox{gray}{yellow}{1}}\\[2mm]
        $1$ & 0 & { \fcolorbox{gray}{yellow}{0}} & 1 & ... & { \fcolorbox{gray}{yellow}{1}}\\[2mm]
        $2$ & 1 & { \fcolorbox{gray}{yellow}{0}} & 0 & ... & { \fcolorbox{gray}{yellow}{0}}\\[2mm]
        $3$ & 0 & { \fcolorbox{gray}{yellow}{1}} & 0 & ... & { \fcolorbox{gray}{yellow}{0}}\\[2mm]
        ... & ... & ... & ... & ... & ...\\[2mm]
        $m$ & 0 & { \fcolorbox{gray}{yellow}{1}} & 0 & ... & { \fcolorbox{gray}{yellow}{0}}\\
\end{tabular}}
    \caption{}\label{tab:1a}
    \label{subfig:set-cover-table}
    \end{subtable}%
    \caption{\textbf{(a)} A hypergraph over twelve vertices with five hyperedges. The minimal hitting set is highlighted in white. \textbf{(b)} An alternative tabular representation of the set cover problem.}
    \label{fig:hs}
\end{figure}
The set cover problem asks, given a set $X$, and a selection $\{V_i\}_{i \in I}$ of subsets of $X$, such that these subsets cover $X$, or $\bigcup_{i \in I} V_i = X$, what is the minimal subset $R$ of the sets $V_i$ which covers $X$? It can be visualised in tabular form, as shown in Fig.~\ref{subfig:set-cover-table}, where the problem becomes how best to choose a set of columns such that every row has at least one `$1$' appearing in one of the columns.  

We solve this problem by instead solving the hitting set problem, since the hitting set and set cover problems can be seen to be equivalent by exchanging the ``is an element of" relation with ``is contained in". More formally, given a set $Y$ of vertices, and a collection $\mathcal{H}$ of subsets of $Y$, the hitting set is a subset $Z \subseteq Y$ such that every $R \in \mathcal{H}$ has at least one element in $Z$. We will refer to this collection $\mathcal{H}$ as a hypergraph. The size of the largest set in $\mathcal{H}$ is the \textit{dimension} of the hypergraph, and its \textit{size} is the number of vertices. An example of a hypergraph is shown in Fig.~\ref{subfig:hypergraph}. It consists of twelve vertices and five hyperedges, with the dimension of each edge indicated by the colour of that edge. The minimum hitting set is highlighted in white. The two problems shown in Fig.~\ref{fig:hs} are equivalent.

An overview of algorithms used to approximate solutions to set cover or the hitting set, and an analysis of their performance, can be found in Ref.~\cite{Caprara_2000}. An important result was reported by Lovász who introduced an algorithm which can find a cover with at most a factor of $1+ \ln d$ vertices greater the optimal solution, where $d$ is the maximum degree of the hypergraph. In terms of solving the problem with energy-based solvers, a QUBO formulation was introduced in Ref.~\cite{Lucas_2014}, and the related problem of set cover with pairs was solved for $m=17$ using quantum annealing and compared against SA, where no quantum speed-up was observed over classical runtimes when averaging over different instances \cite{Cao_2016}. Others have used quantum annealing and neuromorphic computing to study vertex cover, which is a specified case of the hitting set when every edge has dimension $2$ \cite{Pelofske_2019, Corder_2018}, where the hypergraph is simply a graph. 

A higher-order binary optimisation problem can be obtained by generalising the energy function for vertex cover, giving the following:
\begin{equation}
    E(\mathbf{s}) = A \sum_{r \in \mathcal{R}} \prod_{v \in r} \left(1 - s_{v} \right) + B \sum_{v \in V(G)} s_v,
    \label{eqn:energy-hs}
\end{equation}
where $s_v$ is the state of the  p-bit at vertex $v$ and $\mathbf{s} = (s_v)_{v \in Y}$ is a choice of state for each vertex in the hypergraph $\mathcal{R}$. The hyperparameters $A$ and $B$ control the contribution of the two terms, and it can be shown that $A>B$ must be taken to ensure that the solution found is a valid cover. Note that Eqn.~\ref{eqn:energy-hs} can be put in the form of Eqn.~\ref{eqn:energy-higher-order} by grouping together terms of each order and determining the elements of the $J^{k}$ tensors for the different orders $k$. Also note that for a hypergraph of dimension $k$, the maximum product occurring in the energy is of order $k$, making the energy a polynomial of order $k$. We show the formulation of Eqn.~\ref{eqn:energy-hs} explicitly in App.~\ref{app:HS-Hamiltonian}.

In App.~\ref{app:HS-UpdateDrive} we show that p-bits' update drive for the hitting set problem is given by: 
\begin{align}
    I_k & = A \sum_{\substack{r \in \mathcal{R} \vert \\ k \in r}} \prod_{\substack{v \in r \vert \\ v \neq k}} \left(1 - s_{v} \right) - B.
    \label{eqn:bias-calculation-HS}
\end{align}

For each p-bit, this scales as at most as $\mathcal{O}(m(k-1))$, where $m$ is the number of hyperedges in $\mathcal{H}$ and $k$ is the dimension of $\mathcal{H}$. In practice, this computation is often significantly more efficient, as the product can be truncated to zero immediately upon encountering any zero variable

\begin{figure}
    \begin{subfigure}[b]{.5\linewidth}\centering
    \includegraphics[width = \textwidth]{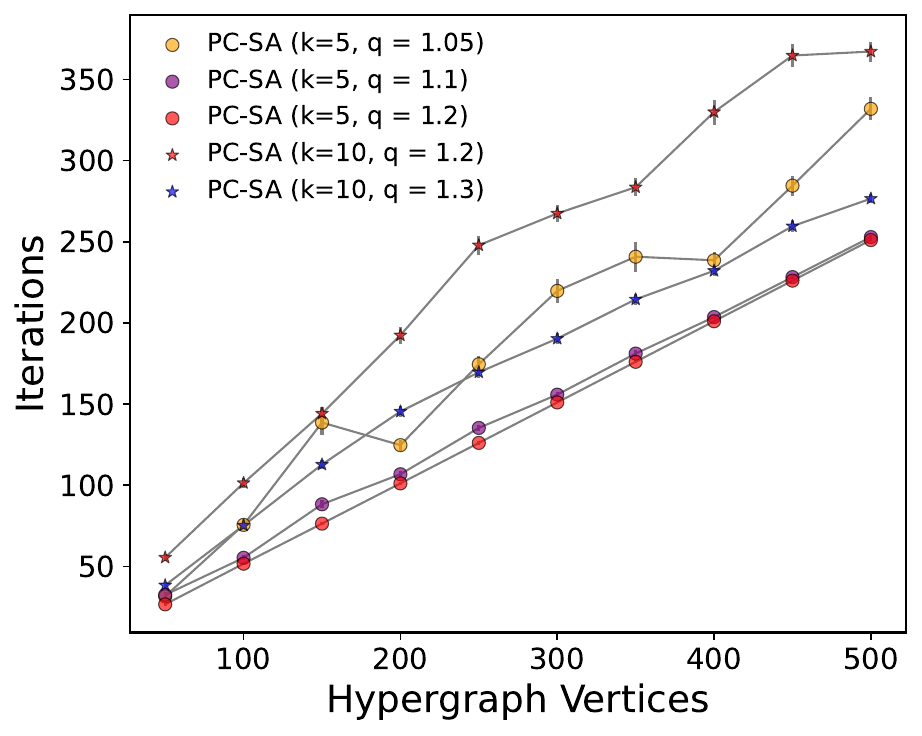}
    \caption{}
    \label{subfig:HS_scaling}
    \end{subfigure} 
    \begin{subfigure}{.5 \linewidth}
    \centering
        \includegraphics[width= \textwidth]{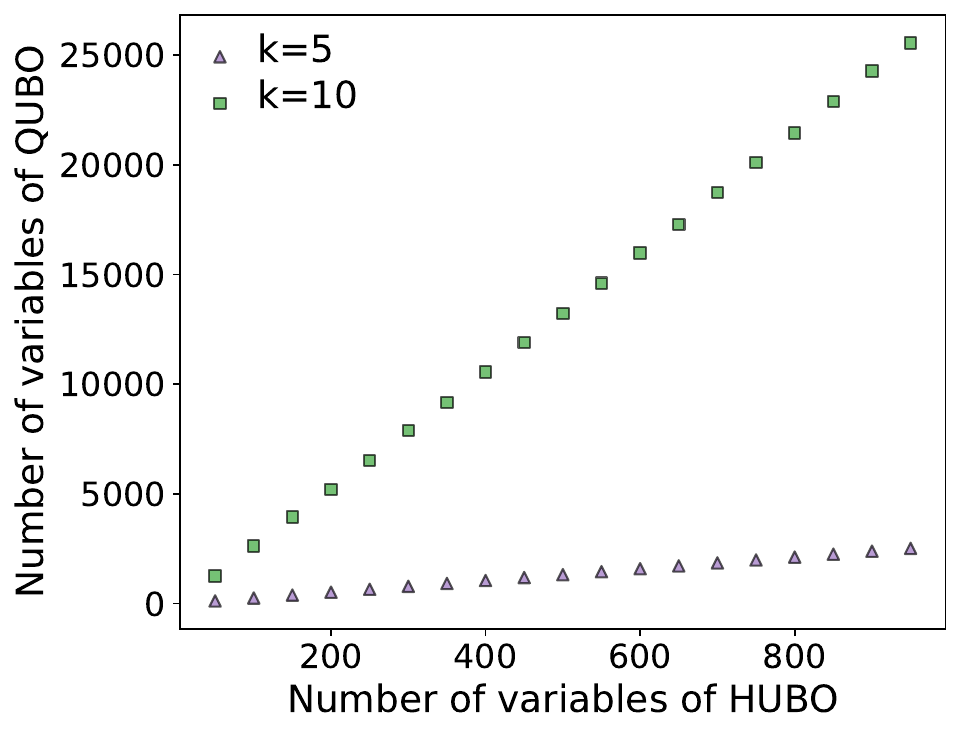}
        \caption{}
        \label{subfig:HS_verts_comparison}
    \end{subfigure}
    \begin{subfigure}{.5 \linewidth}
        \centering
        \includegraphics[width= \textwidth]{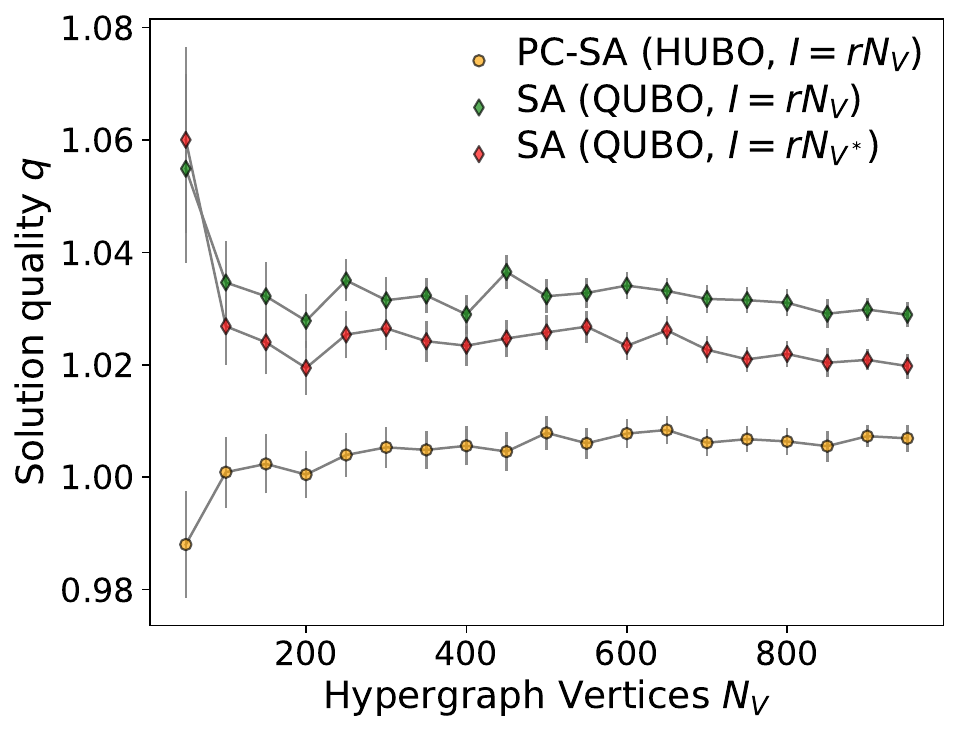}
        \caption{}
        \label{subfig:HS_QUBO-HUBO-quality}
    \end{subfigure}
    \caption{\textbf{(a)} The number of iterations required to reach certain solution qualities for hypergraph dimensions $k=5$ and $k=10$, where now a single iteration updates a group of independent p-bits in parallel. \textbf{(b)} The growth in problem size for randomly generated instances for hypergraph dimensions $k=5$ and $k=10$, when quadratising the HUBO into a QUBO. \textbf{(c)} The performance of the PC-SA algorithm, as measured by solution quality, against a standard QUBO SA solver, where the cooling schedules are equivalently scaled (see text).}
    \label{fig:HS-results}
\end{figure}

Hypergraph dimensions of $k=2$ reduce naturally to QUBO form, so in order to test the performance of the PC-SA and PC-PT algorithms on higher-order problems we require that $k>2$. We conducted our initial performance tests using $k=5$ and examine the impact of increasing $k$ on algorithmic performance in App.~\ref{app:HS-dimension}. We generated $100$ instances for different choices of size in the range of $50$ vertices to $1000$ vertices. The results for $k=5$ and $k=10$ are shown in Fig.~\ref{subfig:HS_scaling}. The hyperparameters used are listed in App.~\ref{subsec:app-hs-hyperparams}. Figure~\ref{fig:HS-results} shows the solution quality $q$, where for this example $q$ is defined as:
\begin{equation}
    q = \frac{\text{HS}_{\mathsf{A}}}{\text{HS}_{\mathsf{ref}}},
    \label{eqn:soln-quality}
\end{equation}
where $\text{HS}_{\mathsf{A}}$ refers to the size of the hitting set found by algorithm $\text{A}$. Usually, we expect $q\geq 1$ as the algorithm will return a value larger than or equal to that of the reference algorithm, although, $q<1$ is also possible here since the reference size was calculated using \texttt{SetCoverPy}~\cite{Zhu2016ANV}, which is itself a heuristic algorithm. 

Figure~\ref{subfig:HS_scaling} depicts the average iterations required to reach target solution qualities for $k=5$ and $k=10$ using a greedy hypergraph colouring algorithm for parallelisation. For the $k=5$ data, the iteration requirement increases only modestly when shifting from $20\%$ to $10\%$ above the reference solution. In contrast, reaching the $5\%$ quality threshold requires a significant jump in the number of iterations, likely reflecting the heuristic nature of the simulated annealing algorithm in finding high-accuracy solutions. For the $k=10$ case, the increased complexity of the energy landscape demands significantly more iterations to achieve a comparable improvement in solution quality. Specifically, shifting from $30\%$ to $20\%$ above the reference necessitates a much larger increase in the iteration count than what was required for the $k=5$ instance.

The general trend shown in Fig.~\ref{subfig:HS_scaling} is that the number of iterations required to reach a particular solution quality appears to scale linearly with vertex size for the ranges investigated. The small number of iterations required is a reflection of the large reduction due to hypergraph colouring. In App.~\ref{app:heat-plots} we investigate how the number of groups, and therefore the speed-up factor, changes with the dimension of the hypergraph and number of edges. We find that for lower hypergraph dimension the best speed-ups are obtained, regardless of the number of edges. On the other hand, for very large dimensions, greedy algorithms perform well, which makes such parameter space less interesting. While group updates reduce the number of iterations required to reach a particular solution quality, we also expect, considering the bias shown in Eqn.~\ref{eqn:bias-calculation-HS}, that the number of clock cycles required to
perform a single iteration, or group update, will depend on $m$ and $k$. A careful analysis of which regimes the speed-up due to group updates undercuts the increase in time needed to compute the bias will be left to future work.

While VCPC is able to work directly with HUBO energy functions, other hardware platforms with physically connected p-bits necessitate quadratisation. To this end, we convert the test instances considered for $k=5$ and $k=10$ to quadratic form, using standard quadratisation procedures, and investigate how (i) the performance difference, for an equal number of iterations, and (ii) the scaling of this conversion process. In Fig.~\ref{subfig:HS_verts_comparison}, we show the growth of the problem size with the quadratisation process for dimension size $k=5$ and $k=10$. This growth for $k=5$ is already quite significant, with the number of QUBO variables corresponding to more than double the number of HUBO variables. For $k=10$, the rate of this growth increases drastically. In particular, the largest problem sizes of $1000$ vertices for the HUBO problem were converted to, on average, $25000$ vertices for the QUBO. 

Figure~\ref{subfig:HS_QUBO-HUBO-quality} shows the performance of the PC-SA algorithm, against a PC-SA solver using the QUBO representation. For each, we have the same starting and ending temperature in the cooling schedule. The strength parameter, which must necessarily be chosen when quadratising an energy function, was optimised for the best performance. The iterations for the PC-SA calculation were taken as $I=rN_V$, where $N_V$ is the number of hypergraph vertices of the HUBO and as detailed in App.~\ref{app:hyperparams}, and we set $r=5$. For the QUBO forms, we performed the calculation with $I=rN_V$, as for the HUBO case, as well as $I=rN_{V^*}$. Here $N_{V^*}$ is the number of vertices of the QUBO after quadratisation (see Fig.~\ref{subfig:HS_verts_comparison}), and as such the number of iterations is significantly higher. As shown in Fig.~\ref{subfig:HS_QUBO-HUBO-quality}, the solution quality for both formulations remains stable as the number of vertices increases, though we note that the iteration count scales linearly with problem size to maintain this performance. Despite this scaling, the QUBO algorithm consistently underperforms the HUBO approach. Crucially, this performance gap persists even when the iteration count for the QUBO is increased to account for the larger number of variables introduced by quadratisation. It is likely that the quadratisation process induces a more rugged energy landscape, as noted in previous studies~\cite{Dobrynin_2024}. This increased complexity explains why simply scaling the iteration count does not allow the QUBO formulation to match the solution quality achieved by the native HUBO.

\subsection{Travelling salesperson problem}

The TSP is an established NP-hard combinatorial optimisation task. Solving the TSP involves finding the minimum tour between $N$ cities $\{A, B, C, ... \}$, such that each city is visited exactly once while ensuring that one returns to the starting city at the end of the tour. It is convenient to define a distance matrix, $D_{ij}$, showing the associated cost of travelling between cities $i \rightarrow j$:
\begin{equation}\label{eqn:D_ij}
    D_{ij} = \begin{pmatrix}
        0 & d_{AB} & d_{AC} & \dots\\
        d_{BA} & 0 & d_{BC} & \dots\\
        d_{CA} & d_{CB} & 0 & \dots\\
        \vdots & \vdots & \vdots & \ddots
\end{pmatrix}
\end{equation}
We assume a symmetric problem where $D_{ij}=D_{ji}$ for simplicity. The energy function for the TSP is given by:
\begin{equation}
\begin{aligned}
E(S)
&= A\Bigg[\sum_{i=0}^{N-1}\bigg(\sum_{k=0}^{N-1}S_{ik}-1\bigg)^{2}
   + \sum_{k=0}^{N-1}\bigg(\sum_{i=0}^{N-1}S_{ik}-1\bigg)^{2}\Bigg] \\
&\!+\, B\Bigg[\sum_{i=0}^{N-1}\sum_{j=0}^{N-1}\sum_{k=0}^{N-2}
   D_{ij}\, S_{ik}\, S_{j,k+1}
   + \sum_{i=0}^{N-1}\sum_{j=0}^{N-1} D_{ij}\, S_{i,N-1}\, S_{j,0}\Bigg]
\end{aligned}
\label{eqn:TSP-Hamiltonian}
\end{equation}
where $S$ is an $N \times N$ matrix with rows corresponding to the cities that can be visited, and columns corresponding to the order in which cities are visited\footnote{While before we used $\mathbf{s}$ to refer to the global state, highlighting that it was a vector, in this context since it is a matrix we use the conventional upper case letter.}. We explain how the ground state of Eqn~\ref{eqn:TSP-Hamiltonian} encodes solutions to the TSP in App.~\ref{app:TSP-Hamiltonian}, and provide the update drive derivation in App.~\ref{app:TSP-UpdateDrive}. The difficulty in solving the TSP lies in the scaling of the number of possible tours, which is well known to be $(N-1)!/2$ for a symmetric problem. Consequently, the brute-force approach becomes unfeasible for problems involving $N>15$ cities, making alternative methods necessary to solve TSP problems within a reasonable timeframe.

One approach is to use KMC to reduce the number of p-bits needed to solve a TSP instance~\cite{Dan_2020, Jaradat_2019}. KMC is a shallow unsupervised machine learning algorithm which partitions data points into $K$ groups. In the context of solving the TSP, we pre-process the problem by recursively clustering the city coordinates $m$ times. Here, clustering refers to a process by which the centroid position of a group of coordinates, which may be cities or clusters of cities, is computed and used as the coordinate in the next round, thereby coarsening the data. Letting $K_i$ denote the $i$th cluster of cities, where $K_1$ represents the original problem, the clustering is performed as $K_{1}\rightarrow K_{2} \rightarrow \dots \rightarrow K_{m}$ where $K_{1} > K_{2} > \dots > K_{m}$, meaning that we are simplifying the problem at each step. 

\begin{figure}[h!]
    \centering
    \begin{subfigure}{.4\linewidth}\centering
    \includegraphics[width = 0.8 \textwidth]{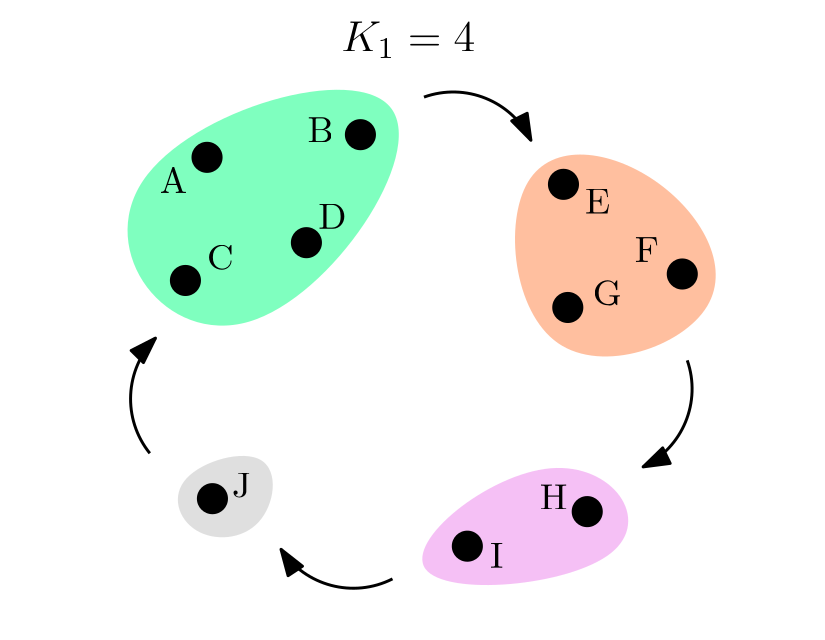}
    \caption{}
    \end{subfigure}
    \begin{subfigure}{.4\linewidth}\centering
    {
$\begin{pmatrix}
1 & 1 & 1 & 1 & 0 & 0 & 0 & 0 & 0 & 0 \\
1 & 1 & 1 & 1 & 0 & 0 & 0 & 0 & 0 & 0 \\
1 & 1 & 1 & 1 & 0 & 0 & 0 & 0 & 0 & 0 \\
1 & 1 & 1 & 1 & 0 & 0 & 0 & 0 & 0 & 0 \\
0 & 0 & 0 & 0 & 1 & 1 & 1 & 0 & 0 & 0 \\
0 & 0 & 0 & 0 & 1 & 1 & 1 & 0 & 0 & 0 \\
0 & 0 & 0 & 0 & 1 & 1 & 1 & 0 & 0 & 0 \\
0 & 0 & 0 & 0 & 0 & 0 & 0 & 1 & 1 & 0 \\
0 & 0 & 0 & 0 & 0 & 0 & 0 & 1 & 1 & 0 \\
0 & 0 & 0 & 0 & 0 & 0 & 0 & 0 & 0 & 1
\end{pmatrix}$}
    \caption{}
    \end{subfigure}
    \caption{\textbf{(a)} Illustration of a single KMC process on an arbitrary 10 city TSP. \textbf{(b)} The mask, $M_{0}$, generated from solving the $K_{1}$ problem, showing how KMC has reduced the number of p-bits required from 100 $\rightarrow$ 29~\cite{Dan_2020}.}\label{fig:tsp_fig1}
\end{figure}

Coarsening the problem to a TSP with $K_{m}$ centroids (where typically $K_{m} \sim 4$) enables the solver to easily find the minimum tour of the reduced problem. This result allows one to construct a mask matrix, $M_{m-1}$, for the $K_{m-1}$ centroid problem, which is a binary matrix of shape $K_{m-1} \times K_{m-1}$. This procedure is depicted in Fig.~\ref{fig:tsp_fig1}. The purpose of $M$ is to dictate which p-bits in $S$ need to be modelled, where 1s indicate that the p-bit should be included, while 0s correspond to clamping the specified p-bit to the 0 state.

The existence of $M$ can be justified by assuming that the optimal tour for the $K_{m-1}$ problem will not comprise moves \textit{between} different clusters until \textit{all} constituent cities have been visited in a given cluster. Ignoring suboptimal routes using this method vastly reduces the search space of the $K_{m-1}$ problem, allowing better approximate solutions. This process is then iterated until we arrive at the original $N$ city TSP, $K_{1}$, where we use the mask generated from solving the $K_{2}$ problem to return a final tour. 

In addition to reducing the number of p-bits, the KMC pre-processing also reduces the number of colour groups needed to cover $S$. For example, using KMC on the 10 city problem shown in Fig.~\ref{fig:tsp_fig1} reduces the number of colour groups from 22 $\rightarrow$ 8. This reduction is due to a combination of using fewer total p-bits as well as fewer conflicts in the remaining p-bits' update drives. Reducing the number of colour groups is significant, as it allows a greater proportion of the global system to be updated in parallel.

The reconfigurable geometry of the VCPC is essential for this process to be implemented on a probabilistic annealer. As each iteration of KMC produces a different set of city groupings and corresponding connectivity patterns, the solver must dynamically adjust which couplings are active. Annealing hardware with fixed physical connections cannot accommodate these changing geometries, whereas the VCPC can reprogram its virtual couplings in real time, enabling adaptation to the evolving problem geometry. The VCPC's reconfigurable geometry enables us to employ KMC in order to improve the solution quality, where both PC-SA and PC-PT are sub-methods used in the KMC procedure. In both cases, we show in Table~\ref{tab:tsp-results}, the KMC pre-processing greatly reduces the average tour cost and enables convergence nearer to the ground state in the best case scenario. It is also worth noting that we are now able to reduce our penalty weight parameter $A$, while maintaining a high fidelity of valid solutions, as there are fewer states contributing to the possible invalid tours available. We show our hyperparameter choices in App.~\ref{subsec:app-tsp-hyperparams}.

\begin{table}[htbp]
    \centering
    \caption{Performances on TSPLIB~\cite{Reinelt_1991} benchmark instances using different solvers. Each solver was run 100 times with different seeds, and the results are shown as a fraction of the known optimal solutions. We do not count invalid tours in these averages, where the number of valid tours per method can be seen in the "Valid" columns.}
    \label{tab:tsp-results}

    \renewcommand{\arraystretch}{1.2}
    \begin{tabularx}{\textwidth}{l
        *{6}{>{\centering\arraybackslash}X}
        *{6}{>{\centering\arraybackslash}X}}
        \toprule
        \multirow{2}{*}{\textbf{Instance}} &
        \multicolumn{6}{c}{\textbf{PC-SA}} &
        \multicolumn{6}{c}{\textbf{PC-PT}} \\
        \cmidrule(lr){2-7} \cmidrule(lr){8-13}
        & \textbf{Best} & \textbf{Ave.} & \textbf{Valid} & \textbf{Best KMC} & \textbf{Ave. KMC} & \textbf{Valid KMC}
        & \textbf{Best} & \textbf{Ave.} & \textbf{Valid} & \textbf{Best KMC} & \textbf{Ave. KMC} & \textbf{Valid KMC}\\
        \midrule
        burma14     & 1.071  & 1.237  & 100 & 1.000 & 1.079 & 99
                    & 1.089  & 1.222  & 100 & 1.000 & 1.082 & 100 \\
        ulysses16   & 1.169 & 1.303 & 100 & 1.007 & 1.072 & 81
                    & 1.117 & 1.238 & 100 & 1.007 & 1.070 & 83 \\
        ulysses22   & 1.306 & 1.559 & 100 & 1.010 & 1.074 & 91
                    & 1.262 & 1.416 & 100 & 1.010 & 1.072 & 91 \\
        berlin52    & 2.221 & 2.543 & 99 & 1.091 & 1.197 & 99
                    & 2.551 & 2.912 & 42 & 1.049 & 1.199 & 99 \\
        \bottomrule
    \end{tabularx}
\end{table}

More specifically, in Table~\ref{tab:tsp-results} we show that for smaller instances: burma14, ulysses16 and ulysses22, KMC enables both SA and PT to converge to within 1$\%$ of the optimal tour for the best case scenario. As well as this, the average tour is consistently $\sim7.5\%$ greater than the optimal tour which represents a vast improvement on the standard PC-SA and PC-PT solvers. For the largest instance considered, berlin52, the KMC procedure approximately halves the best and average tour costs for both PC-SA and PC-PT. Remarkably, the fidelity of valid tours is consistently $>80\%$ for every instance and method investigated with the exception of standard PC-PT on berlin52. Despite this, the addition of the KMC procedure in the PT solver enables a best tour cost of 1.049 to be achieved. Our scores for berlin52 are comparable to digital annealing results as reported in Ref.~\cite{Ayodele_2022}. Two improvements that could be made would be to utilise the recently published 2D-PT~\cite{Delacour_2025} and Transformer-Augumented-PT~\cite{Bunaiyan_2025} algorithms, to ensure constraint satisfaction and propose pre-learned global moves, respectively.

\subsubsection{Sparsification}

It is known that dense graphs are particularly difficult to embed into hardware~\cite{Okada_2019}. Relatively little is reported in the literature related to the geometry of the TSP problem, both with and without KMC. Since embeddings or sparsification must be carried out to implement the problem on hardware, understanding this geometry helps to shed light on the difficulty of doing so.

The density of a graph $G$ with $N_E$ edges ($N_E \approx 2N^{3}$) and $N_V$ vertices ($N_V=N^{2}$) is defined as $M=\frac{N_E}{N_V\times (N_V-1)}$, where $N_V \times (N_V-1)$ is the maximum possible number of edges. As such, the density of the TSP problem can be expressed as:
\begin{equation}
    M \approx \frac{2N^3}{N^4} = \frac{2}{N},
\end{equation}
where $N$ is the number of cities (or equivalently, the number of time steps). Since $M \rightarrow 0$ as $N \rightarrow \infty$, the problem is not strictly dense, but decays more slowly than any other class of non-dense graphs. 

Another geometrical property affecting the embeddability is the size of the largest connected component in the geometry, corresponding to the largest subset of vertices which forms a complete graph. This structure is particularly difficult to embed, and places a lower bound on the growth of the problem when embedding is performed. We find that, for the berlin52 instance with KMC, the biggest connected component is of order $100$ vertices. Therefore, both the high number of connections and large connected components would make these difficult problems to find efficient embeddings for, or to sparsify without a substantial increase in problem size.

One way to circumvent such embedding difficulties is to break the problem down into subproblems. Although, in principle, one could break down the problem at each step in KMC, the geometry of the problem dynamically changes at each step due to the uniqueness of the obtained mask. This means that the embedding must be calculated $m$ times, and cannot be pre-computed, significantly increasing the complexity when running on restricted physically connected hardware. This also applies to sparsification, as the problem must be sparsified for each and every KMC step.

We now consider a graph sparsification procedure recently published in Ref.~\cite{Sajeeb_2025}, in the context of solving the TSP. The purpose of graph sparsification is to reduce the density of a problem by enforcing a limit on each p-bit's number of neighbours, at the cost of introducing additional auxiliary p-bits. This connectivity regulation is crucial for non-virtually connected architectures, where large scale fully connected graphs become unfeasible in hardware. The graph sparsification algorithm introduces copy nodes to limit the connectivity to the hyperparameter $k$. Ferromagnetic edges are then introduced between a source node (the original p-bit) and it's copy nodes in order to correlate their behaviour.

\begin{figure}[h]
\centering
    \begin{subfigure}{0.49 \linewidth}
    \centering
    \includegraphics[width=\linewidth]{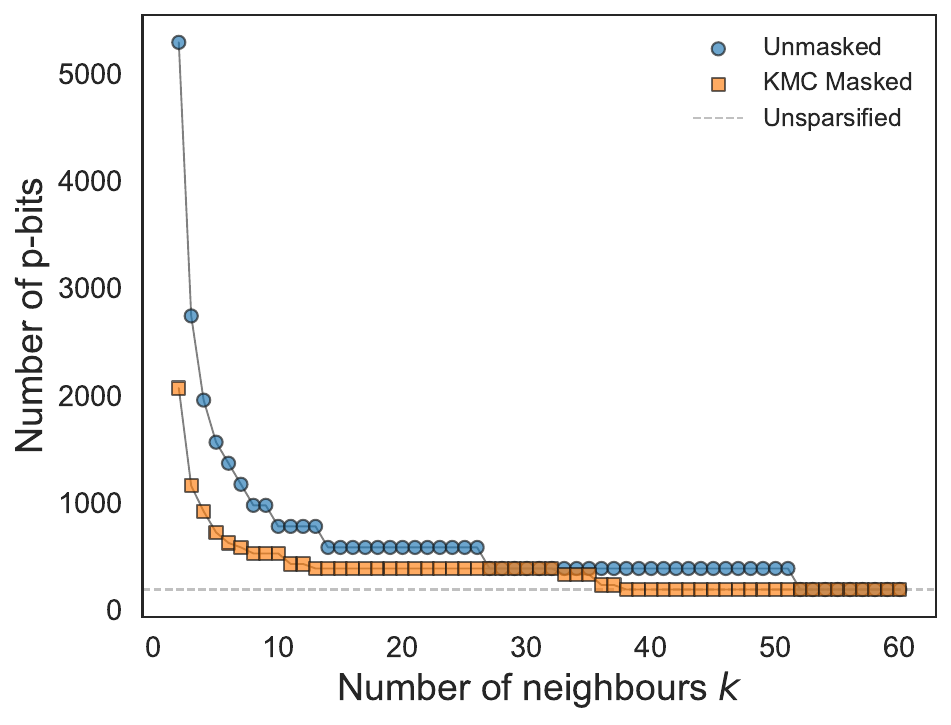}
    \caption{}
    \label{subfig:sparse-burma14-a}
    \end{subfigure}
   \begin{subfigure}{0.49 \linewidth}
   \centering
    \includegraphics[width=\linewidth]{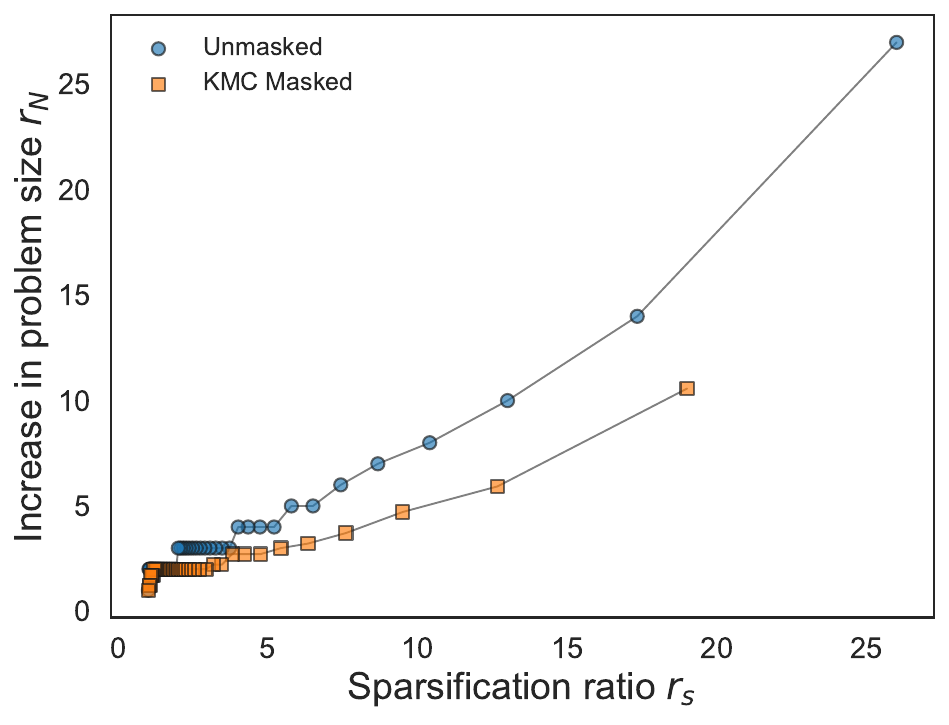}
    \caption{}
    \label{subfig:sparse-burma14-b}
    \end{subfigure}
    \caption{Sparsification results for the burma14 instance. \textbf{(a)} The number of p-bits required to reach an architecture with a specified number of neighbours, with a low neighbour count requiring more p-bits. \textbf{(b)} The relationship between the ratios $r_N$ and $r_S$ representing the increase in problem size and the sparsification of the graph, respectively (for $k=2$). Results are included for both KMC masked and unmasked p-bit arrays~\cite{Sajeeb_2025}.}
    \label{fig:sparse-burma14}
\end{figure}

In Fig.~\ref{subfig:sparse-burma14-a} the relationship between the number of required p-bits and the connectivity is shown. As expected, for a lower neighbour count, the number of p-bits is significantly higher. The unmasked architecture is also more difficult to sparsify, as for equal neighbour counts, the number of p-bits required is always greater than that of the KMC masked version. It can be seen in Fig.~\ref{subfig:sparse-burma14-a}, for example, that constraining each p-bit to have at most $10$ neighbours results in a problem of size of $\sim1000$ p-bits for the unmasked version and $\sim500$ for the masked version, while the original problem size --- indicated by the grey dotted line --- required only $196$ p-bits.

We introduce two parameters to further help visualise the scaling of the sparsification procedure on TSP instances. If $G$ denotes the connectivity of the original TSP problem, and $G_S$ denotes the sparsified graph, then we let $r_N$ be the ratio of the number of vertices in $G_S$ to the number of vertices in $G$:
\begin{equation}
    r_N = \frac{N_V(G_S)}{N_V(G)}.
\end{equation}

We also introduce a sparsification ratio, $r_S$, which shows how the maximum number of neighbours decreases as auxiliary variables are included. Specifically,
\begin{equation}
    r_S = \frac{N(G)}{N(G_S)}
\end{equation}
where $N(G)$ is the maximum number of neighbours in a graph $G$.

In Fig.~\ref{subfig:sparse-burma14-b} we show how the ratio $r_N$ varies, for both the unmasked and KMC masked methods, as the ratio $r_S$ increases. As expected, the KMC masked p-bit array scales more favourably than the unmasked case, as the KMC masked problem is initially denser. For a graph which is $10$ times more sparse than the original, the problem increases in size by a factor of at least $5$. It is also worth noting that both cases start at the point $(1, 1)$, corresponding to the unsparsified case.

\newpage
\subsection{Spin-glass}
\label{subsec:spin-glass}

The final NP-hard problem we consider is the problem of finding low-energy states of a spin-glass lattice, which is defined as a system of $N$ spins with a symmetric $N \times N$ coupling matrix $J$. The elements of the matrix encode ferromagnetic and anti-ferromagnetic couplings between lattice sites, the frustration of which makes the problem complex. Starting from the Ising Hamiltonian with an external field: 
\begin{equation}
    \mathcal{H} = -\sum_{i < j} J_{ij} \sigma_i \sigma_j -\sum_i h_i \sigma_i
    \label{eq:Hamiltonain}
\end{equation}
where $\sigma_{i} \in \{-1,+1\}$, we map this to a QUBO problem by applying the substitution $\sigma_{i}=2s_{i}-1$ where $s_{i} \in \{0,1\}$. This transformation results in the QUBO energy function:
\begin{equation}
    E(\mathbf{s}) = \sum_{i < j} Q_{ij} s_i s_j + \sum_i b_i s_i
\end{equation}
where $Q_{ij}$ (for $i<j$) and $b_{i}$ are the transformed coupling and external bias terms dervied from $J_{ij}$ and $h_{i}$, under the bipolar to binary mapping (shown explicitly in App.~\ref{app:SG-Mapping}). The goal is to find the combination of values of $\mathbf{s}$ that minimise the energy of this lattice according to these connections, or a close approximation of it. This problem can also be phrased in terms of the weighted Max-Cut problem, where one optimises for splitting a graph in two, such that the highest summed weight of edges crosses the divide. We show the p-bit update drive derivation in App.~\ref{app:SG-UpdateDrive}.

Different lattice topologies can be encoded by populating $J$ differently, ranging from the Edwards-Anderson (EA) model~\cite{Mezard_1987, Edwards_1975, Sherrington_1975, Binder_1986} of nearest-neighbour-interactions in a D-dimensional lattice to the Sherrington-Kirkpatrick model of full-connectivity~\cite{Sherrington_1975, Camargo_2021, Panchenko_2012}. In Ref.~\cite{Fredrik_2025} spin-glasses were solved for cubic (sparse) lattices and biclique (dense) lattices with probabilistic computing and for this section we focus more specifically on the effect of graph connectedness on performance.

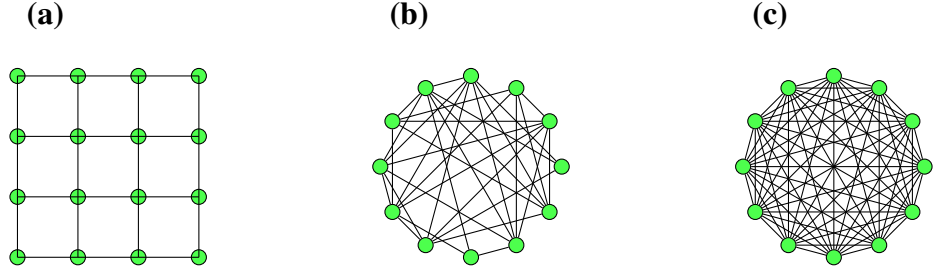
\begin{figure}[ht]
\centering
\begin{tikzpicture}[scale=0.4,
  vertex/.style={circle,draw,fill=green!70,inner sep=2pt},
  edge/.style={}
]

\begin{scope}[shift={(0,0)}]
  \node[anchor=west] at (-3, 5) {\large\textbf{(a)}};
  \def\La{4}              
  \def\dx{2.0}            

\pgfmathsetmacro{\half}{(\La-1)/2.0}
\foreach \ii in {0,...,\numexpr\La-1\relax}{
  \foreach \jj in {0,...,\numexpr\La-1\relax}{
    \coordinate (A\ii\jj) at ({(\ii-\half)*\dx}, {(\jj-\half)*\dx});
    \node[vertex] at (A\ii\jj) {};
  }
}

  \foreach \ii in {0,...,\numexpr\La-1\relax}{
    \foreach \jj in {0,...,\numexpr\La-1\relax}{
      \ifnum\ii<\numexpr\La-1\relax
        \pgfmathtruncatemacro{\iip}{\ii+1}
        \draw[edge] (A\ii\jj) -- (A\iip\jj);
      \fi
      \ifnum\jj<\numexpr\La-1\relax
        \pgfmathtruncatemacro{\jjp}{\jj+1}
        \draw[edge] (A\ii\jj) -- (A\ii\jjp);
      \fi
    }
  }
\end{scope}

\begin{scope}[shift={(12,0)}]
  \node[anchor=west] at (-3, 5) {\large\textbf{(b)}};

  \def\N{12}           
  \def\R{3}            
  \def\pEdge{0.5}      
  \pgfmathsetseed{2024} 

  \foreach \ii in {1,...,\N}{
    \path coordinate (p\ii) at ({\R*cos(360*\ii/\N)}, {\R*sin(360*\ii/\N)});
    \node[vertex] (v\ii) at (p\ii) {};
  }

  \foreach \ii in {1,...,\N}{
    \pgfmathtruncatemacro{\istart}{\ii+1}
    \foreach \jj in {\istart,...,\N}{
      \pgfmathsetmacro{\u}{rnd}
      \ifdim \u pt<\pEdge pt
        \draw[edge] (v\ii) -- (v\jj);
      \fi
    }
  }
\end{scope}

\begin{scope}[shift={(24,0)}]
  \node[anchor=west] at (-3, 5) {\large\textbf{(c)}};

  \def\NSK{12}          
  \def\Rsk{3}           

  \foreach \kk in {1,...,\NSK}{
    \path coordinate (s\kk) at ({\Rsk*cos(360*\kk/\NSK)}, {\Rsk*sin(360*\kk/\NSK)});
    \node[vertex] (w\kk) at (s\kk) {};
  }

  \foreach \kk in {1,...,\numexpr\NSK-1\relax}{ 
    \pgfmathtruncatemacro{\kstart}{\kk+1}
    \foreach \mm in {\kstart,...,\NSK}{
      \draw[edge] (w\kk) -- (w\mm);
    }
  }
\end{scope}
\end{tikzpicture}
\caption{%
    Spin-glass topologies of 
    \textbf{(a)} Edwards–Anderson model on a 2D lattice (sparse). 
    \textbf{(b)} Erdős–Rényi random graph (dense). 
    \textbf{(c)} Sherrington–Kirkpatrick all-to-all model (dense \& maximally connected).%
  }
  \label{fig:spin-glass-topologies}
\end{figure}

To this end, we focus on Erdős–Rényi graphs~\cite{Erdos_1960}, where all possible edges have a probability of $p$ of being included in the graph. We consider a range of $p$-values, taking the graphs through different degrees of connectivity. In the limit of $p\rightarrow1$, the graph becomes fully-connected, whereas the $p=0$ graph is disconnected and its optimal configuration is trivial. In any $p \neq 0$ configuration, the topology of the lattice is dense in that as $N\to\infty$ the ratio of obtained versus possible edges approaches $p$, not zero. Within a given topology, different Max-Cut versions can be formulated depending on  how the values $J_{ij}$ are chosen. For this work we follow Ref.~\cite{Aramon_2019} in considering edges sampled from a set $\{-1,+1\}$ with equal probability.

Finding the update-groups of Erdős–Rényi graphs is itself an NP-hard problem of colouring a random graph. For practical purposes, we use a greedy colouring algorithm. While this approach does not guarantee an optimal colouring, it operates in polynomial time and ensures the use of no more than \(\Delta + 1\) colors, where \(\Delta\) denotes the maximum degree of the graph. This makes it a practical and efficient fallback when exact graph colouring is computationally intractable \cite{Borowiecki_2018}. In 1976 Bollobás \cite{Bollobás_1976} proved that the chromatic number of an Erdős–Rényi graph with constant $p$ and $n$ vertices is with very high likelihood
\begin{equation}
    \chi\bigl(G(n,p)\bigr) = \frac{n}{2 \log n} \log(\frac{1}{1-p}){}\,(1 + o(1)).
\end{equation}
This increases the efficacy of probabilistic computers when either a low probability or a low number of vertices lead to a low number of update groups, with the limit of a bipartite graph being the optimal configuration.

\begin{figure}
    \begin{subfigure}{.5 \linewidth}
\centering
    \includegraphics[width=\textwidth]{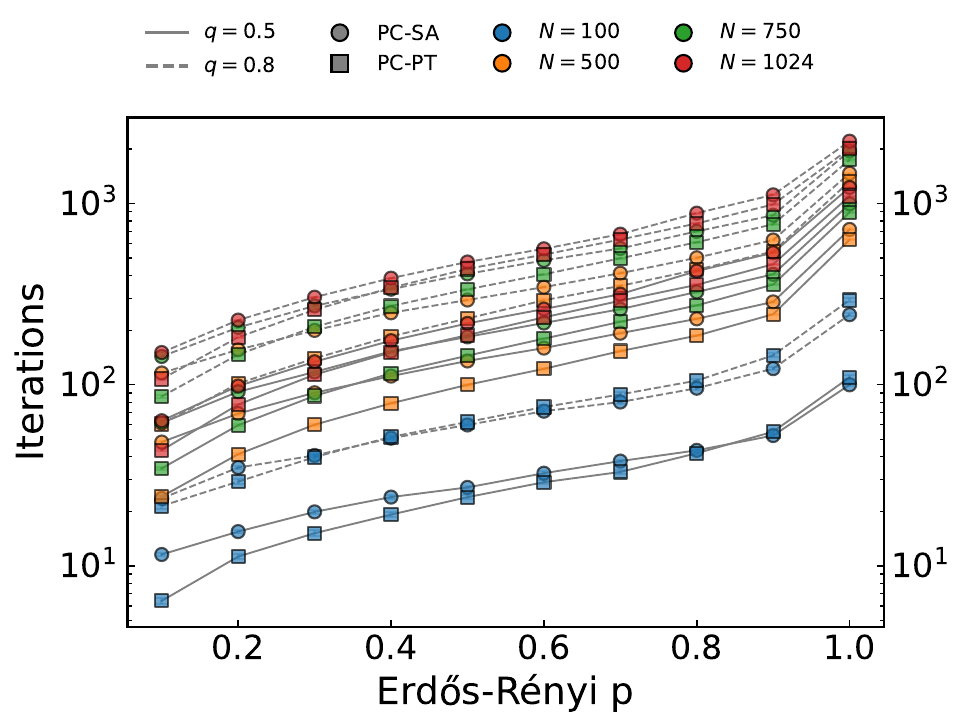}
    \caption{Iteration scaling with density at various sizes}
    \label{subfig:ER-with-p}
\end{subfigure}
    \begin{subfigure}{.5\textwidth}
\centering
\includegraphics[width=\textwidth]{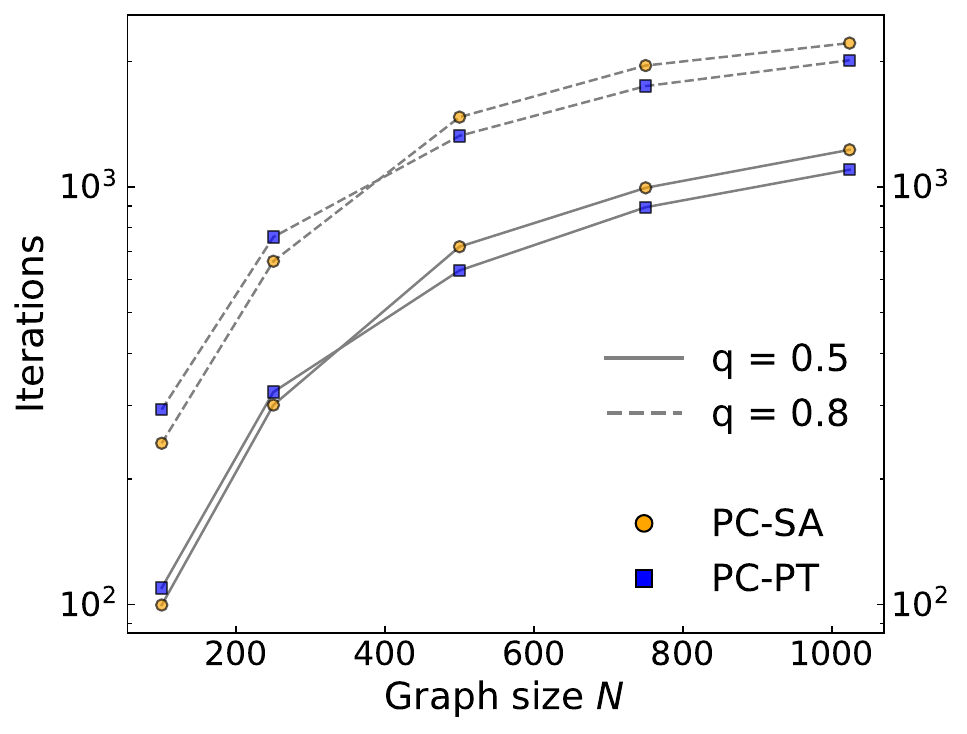}
    \caption{Iteration scaling with graph size at $p=1$}
    \label{subfig:ER-with-N}
\end{subfigure}
    \caption{The number of iterations required to attain solution qualities of $q=0.5,0.8$ over \textbf{(a)} the Erdős–Rényi graph density $p$ and \textbf{(b)} the graph size $N$ at a fixed $p=1$. Note we redacted the $N=250$ data in \textbf{(a)} for better visual clarity.}
    \label{fig:SG-ER-results}
\end{figure}

We consider spin-glasses on Erdős–Rényi graphs for probabilities ranging from $p=0.1$ to $p=1$. We choose solution qualities of $0.5$ and $0.8$ in order to directly compare to results reported in Ref.~\cite{Nikhar_2025}, where SA, PT, and digital annealing were considered. The authors found that the TTS for digital annealing showed a significant improvement over SA and PT algorithms, with the TTS of graph sizes up to $N=1024$ being below 1 second for digital annealing, and growing up to order $10^3$ seconds for parallel tempering.

We therefore consider lattice sizes $N= 100, 250, 500, 750, 1024$ and take $q=0.5$ and $q=0.8$, computing the number of iterations required to reach such qualities\footnote{Note that in Ref.~\cite{Nikhar_2025} they use the terminology `cutoff' rather than solution quality, but the two are equivalent.} shown in Fig.~\ref{subfig:ER-with-p}. In Fig.~\ref{subfig:ER-with-N}, we show how the number of iterations scales with increasing $N$ when the probability of including an edge is fixed at $p=1$. We approximate the ground states using large PC-PT simulations (see App.~\ref{subsec:app-sg-hyperparams} for the number of iterations and replicas used).

Note that for SA, the number of iterations is the total number $\mathcal{I}$, and in particular we take $I=1$, so that at each $\beta$-step only one iteration is performed. On the other hand, for PC-PT the number of replicas is not accounted for in the iteration count. This is because replicas can in theory be run in parallel in any hardware implementation. Based on the number of iterations and placing only moderate assumptions on hardware, we can make estimates from Fig.~\ref{fig:SG-ER-results} about the TTS we expect when running such simulations on dedicated probabilistic hardware.

\subsubsection{Time to solution estimation}\label{subsubsec:SG-TTS}
\begin{figure}[h]
\centering
\includegraphics[width=0.6\linewidth]{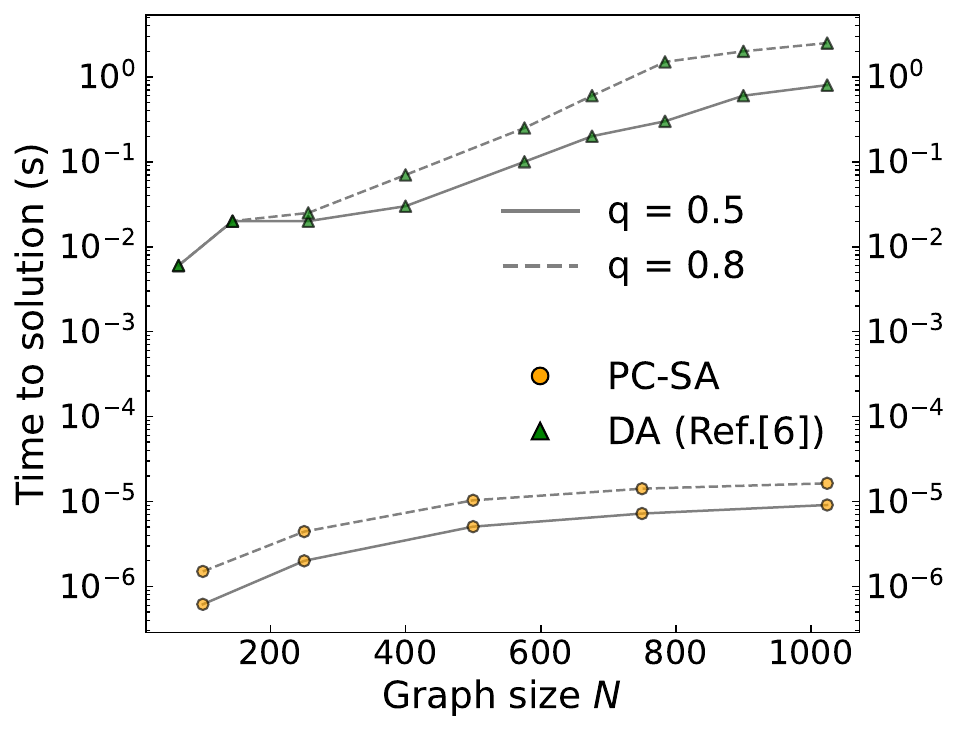}
\caption{Comparison between TTS estimates for PC-SA using Eqn.~\ref{eqn:tts-est} and reported digital annealing values from Ref.~\cite{Aramon_2019}, for Erdős–Rényi graphs with $p=1$.}
\label{fig:SA-vs-Da}
\end{figure}

Recall from Eqn.~\ref{eqn:complexity_group2} that the relationship for estimating the scaling of operations to reach a solution of quality $q$ with probability $p$ for a particular problem is:
\begin{equation}
    O_{q,p} = \frac{\mathcal{I}_{q, p} \cdot \mathcal{O}(p(\mathbf{n}))}{\bar{G}},
\end{equation}
assuming all $R$ repeats can be done in parallel. We can similarly get an estimate for the TTS if, rather than computing the number of operations needed to perform the update drive calculation once, we compute the number of clock cycles required by the hardware to compute the update drive, which we will denote $n_{\text{cycles}}(\mathbf{n})$. Then, given that the hardware runs at a frequency $f$, which is to say in one second it performs $f$ cycles, the total TTS is: 
\begin{equation}
    \text{TTS}_{q,p} = \frac{\mathcal{I}_{q,p} \cdot f^{-1} \cdot n_{\text{cycles}}({\mathbf{n}})}{\bar{G}}
\label{eqn:TTS-scaling}
\end{equation}
where recall that $\mathbf{n}$ represents problem parameters, which for spin-glass would be the size of the problem, the density of edges, and so forth. Given an equation for computing the update drive, estimates can be made about the scaling of $n_{\text{cycles}}(\mathbf{n})$. We compute that the number of clock cycles required to do a single update drive calculation is at most:
\begin{equation}
    n_{\text{cycles}} = \log_2(N) + \mathcal{O}(10),
\end{equation}
where the first term comes from the adder tree algorithm for computing multiplications, and the second term is a constant overhead accounting for miscellaneous background processes. Taking a clock speed of $\SI{2.7}{\GHz}$, which was provided in Ref.~\cite{Ramy_2025}, Eqn.~\ref{eqn:TTS-scaling} gives: 
\begin{equation}\label{eqn:tts-est}
    \text{TTS}_{p,q} = \frac{1}{\bar{G}} \cdot \mathcal{I}_{p,q} \cdot \frac{1}{2.7 \times 10^9} \cdot \left[\log_2(N) + \mathcal{O}(10) \right].
\end{equation}
The factor of $\frac{1}{\bar{G}} \cdot \mathcal{I}_{p,q} $ for the largest lattice size of $N=1024$ can be read off of Fig.~\ref{subfig:ER-with-N}, and is approximately $2000$ (note that group updates were used in calculating the number of iterations). Although the number of repeats do not in theory affect this scaling, since they can be run in parallel, it is important to note hardware limitations. However, here we have only used a single repeat and therefore need not consider hardware limitations in this respect. As shown in Fig.~\ref{fig:SA-vs-Da}, we estimate the TTS for $q=0.8$ as $\sim10^{-5}$ seconds for the largest graph size of $1024$, outperforming the digital annealing algorithm considered in Ref.~\cite{Aramon_2019} by several orders of magnitude. 

\subsubsection{Sparsification}\label{subsubsec:SG-Sparsification}

Finally, we consider the sparsification of spin-glasses. For Erdős–Rényi random graphs, the density $M$ is given by:
\begin{equation}
    M = \frac{p N(N-1) }{N(N-1)} = p.
\end{equation}
Consequently, when $p=1$ these graphs are fully connected, with each node having $N-1$ neighbours. The difficulty of embedding such graphs into hardware is well-documented~\cite{Okada_2019}. Using efficient embedding algorithms on D-Wave’s 2000-qubit system, the maximum embeddable problem size scales with density: it is approximately 275 variables for sparse graphs ($p=0.01$) and decreases to 60 variables for complete graphs ($p=1$).

\begin{figure}[h]
\centering
\begin{subfigure}{0.49 \textwidth}
\centering
        \includegraphics[width=\linewidth]{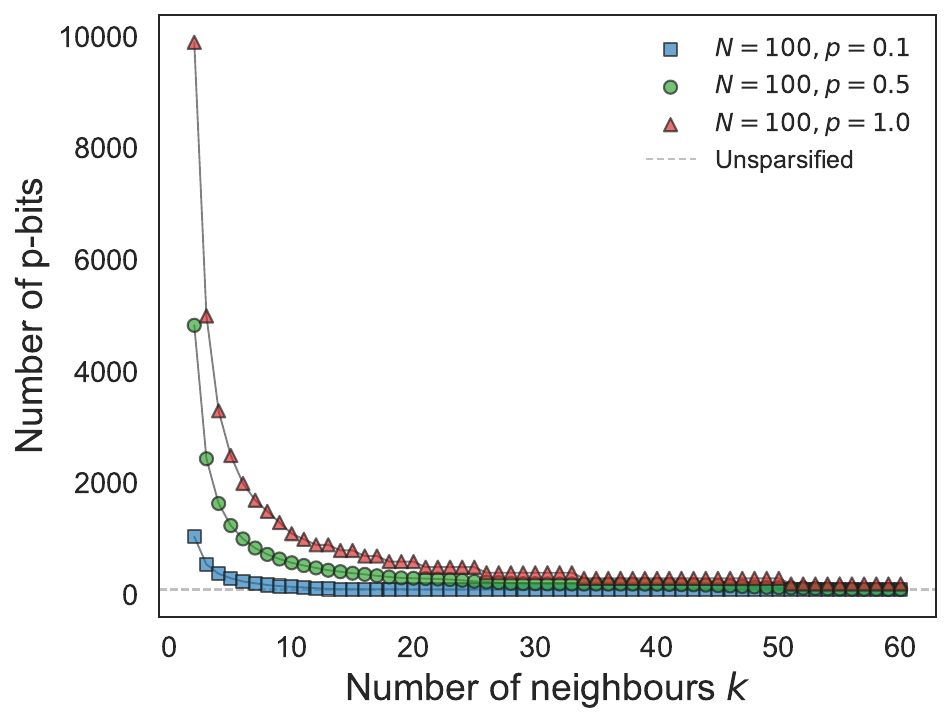}
    \caption{}
    \label{subfig:sparse-SG-a}
\end{subfigure}
\begin{subfigure}{0.49 \textwidth}
\centering
    \includegraphics[width=\linewidth]{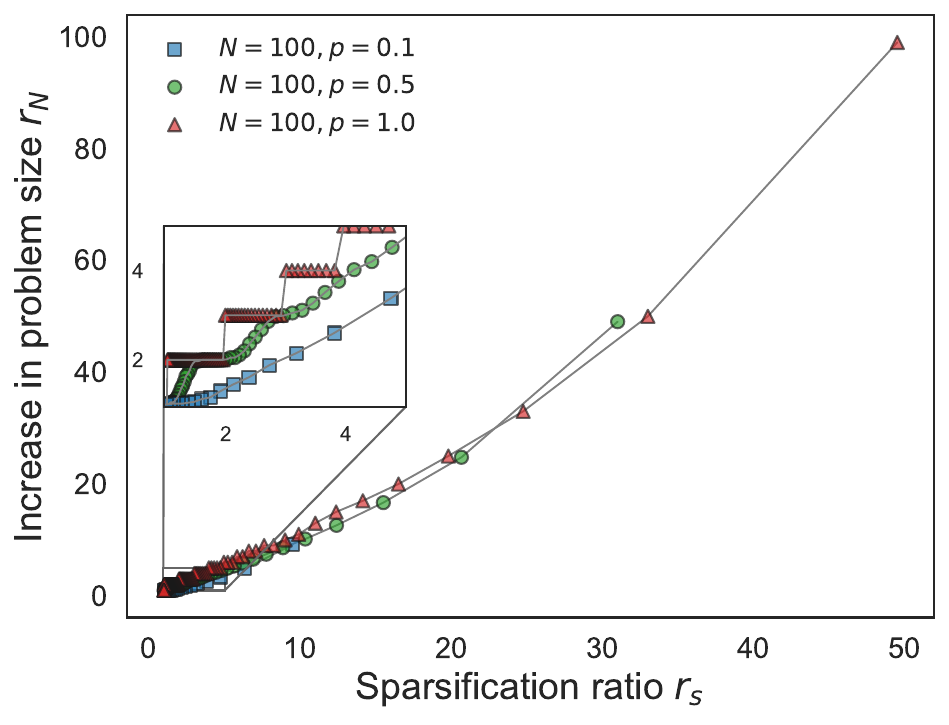}
    \caption{}
    \label{subfig:sparse-SG-b}
\end{subfigure}
    \caption{The sparsification results for an Erdős–Rényi spin-glass instance with $N=100$. \textbf{(a)} The number of p-bits required for a specified maximum number of neighbours for the sparsified graph \textbf{(b)} shows the relationship between the ratios $r_N$ and $r_S$ representing the increase in problem size and the sparsification of the graph (for $k=2$).}
\label{fig:sparse-SG}
\end{figure}

Here, as with the TSP, we focus on sparsification results since these are hardware independent. In Fig.~\ref{fig:sparse-SG} we show the effect of sparsification on the graph size for an $N=100$ test instance, with $p=0.1, 0.5, 1$. Figure~\ref{subfig:sparse-SG-a} shows the number of p-bits needed for a specified number of neighbours in the sparsified graph. As the number of allowed neighbours increases, the number of p-bits in the unsparsified problem is eventually recovered. For the $p=1$ instance, restricting the number of neighbours to less than $10$, requires over $1000$ to implement the sparsified problem. Many hardware implementations would suffer subject to these constraints, including the Chimera graph structure used by D-Wave Inc., where each qubit is limited to $6$ neighbours. Figure~\ref{subfig:sparse-SG-b} shows how the ratio $r_N$, representing the increase in problem size compared to the unsparsified version, increases as the sparsification ratio $r_s$ increases. Again, for $p=1$, the growth is most significant: in the worst case, sparsifying the original graph by a factor of $50$ results in an increase in the size of the problem by a factor of $100$.

%% file: Contents/Conclusion.tex
\newpage
\section{Conclusion}

\begin{table}[]
    \centering
    \begin{tabular}{|c|cccc|}
        \hline
        & PC & DA & QA & VCPC \\
        \hline \hline
         Dense & \ding{55} & \ding{51} & \ding{55} & \ding{51} \\
         Reconfigurable & \ding{55} & \ding{51} & \ding{55} & \ding{51} \\
         Higher-order & \ding{55} & \ding{55} & \ding{55} & \ding{51} \\
         \hline
    \end{tabular}
    \caption{A comparison of the ability of different Nature inspired computing platforms to implement the types of problem considered in this study, without resorting to quadratisation, embedding, or sparsification. We compare the VCPC with (non virtually connected) probabilistic computing (PC), digital annealing (DA), and quantum annealing (QA) platforms.}
    \label{tab:hardware-comparison}
\end{table}

We have presented an argument for the use of a VCPC on problems with geometries that are difficult to implement on hardware lacking virtual connections. 

In Section~\ref{sec:problems}, we demonstrated the performances of the PC-SA and PC-PT algorithms on NP-hard problems by conducting VCPC simulations. For the hitting set problem, we showed that the VCPC achieved solutions qualities consistently within 1\% of the reference solution for hypergraphs up to 1000 vertices (for dimension $k=5$). Crucially, this performance relies on greedy hypergraph colouring to facilitate parallel updates. We found that for hypergraph dimensions $k \leq 6$, the number of update groups is typically less than 10, providing an order of magnitude speed-up for many instances.

For the TSP, we showed that the addition of KMC significantly improved the solution quality. For the 52-city problem (berlin52), the KMC procedure reduced the best and average tour costs be $\sim50\%$ compared to standard PC-SA and PC-PT, with a valid tour fidelity across all instances exceeding $80\%$. In our study of the spin-glass problem, we estimated the TTS for Erdős–Rényi graphs up to $N=1024$. Using a 2.7 GHz clock frequency, we estimate that the VCPC could achieve a TTS of order $10^{-5}$ seconds, outperforming the digital annealing results reported in Ref.~\cite{Aramon_2019} by orders of magnitude.

In Table \ref{tab:hardware-comparison}, we emphasise these results by highlighting the advantages of VCPC over other platforms. While fully connected digital annealers are capable of implementing any QUBO, they necessitate quadratisation, which we showed leads to a problem size increase by a factor of 25. Also, in our hitting set study, we showed that QUBO formulations struggle to reach the same solution quality as the native HUBO formulation. Furthermore, both quantum annealers and physically connected probabilistic architectures suffer from server scaling limitations due to the overhead of embedding and sparsification. By circumventing these pre-processing requirements and enabling reconfigurable, all-to-all virtual connectivity, the VCPC provides a scalable solution for complex, higher-order, and dynamically changing combinatorial optimisation problems.

%% file: Contents/Contributions.tex
\section*{Author Contributions}
\label{Contributions}
A.J.S. conceptualised the project. A.J.S. and H.Y. performed simulations and authored the manuscript, while A.J.S., H.Y., and F.H. wrote python implementations. A.M. oversaw hardware latency estimates, and M.L. and R.A. supervised and secured funding for the project. All authors contributed to editing of the manuscript and substantially to discussions on the results and ideas contained therein. 

%% file: Contents/Acknowledgments.tex
\section*{Acknowledgments}
\label{sec:acknowledgments}
This work was supported by the European Innovation Council's (EIC) grant number 101248376 (Q-TASTic).

%% file: Contents/Appendices.tex
\newpage
\appendix
\section{Hyperparameters}
\label{app:hyperparams}

We indicate which hyperparameters were used for each problem, to ensure reproducibility of the results. When the number of repeats is not specified, it was set to $1$.

\subsection{Hitting set problem}
\label{subsec:app-hs-hyperparams}
For the hitting set calculations in Fig.~\ref{fig:HS-results}, we use the hyperparameters indicated in Table \ref{tab:parameters-HS-PT-SA}, where $N$ is the number of vertices in the hypergraph. In Fig.~\ref{subfig:HS_scaling}, iterations refers to the product $\mathcal{I} = I \times S$, and we keep $I$ fixed at one and only vary $S$. Thus,  $\mathcal{I} = S$. All other parameters are as in Table~\ref{tab:parameters-HS-PT-SA}. In all experiments we took $A=13$ and $B=9$. 

\begin{table}[htbp]
    \centering
    \caption{The hyperparameters used for hitting set instances.}
    \begin{tabular}{ccccccc}
    \toprule
        & \textbf{Iters.}  & \textbf{Steps}  & \textbf{Reps.} & \textbf{Repls.} & \textbf{Temp. range} & \textbf{Swap} \\
        \midrule
        \textbf{SA} & $5 \times N$ & $100$ & $20$ & - & $0.01-1.1$ & -\\
        \textbf{PT} & $50 \times N$ & - & $10$ & $20$ & $0.5-10$ & $25$ \\
        \bottomrule
    \end{tabular}
    \label{tab:parameters-HS-PT-SA}
\end{table}

\subsection{Travelling salesperson problem}
\label{subsec:app-tsp-hyperparams}
For the TSP, hyperparameters are displayed in Table \ref{tab:tsp-hparams}. $K_{m}$ denotes the number of clusters used at each level and $A_{m}$ is the corresponding penalty weight used in the solver. $A_0$ represents the penalty weight for the final $N$-city problem. The penalty weight for the SA and PT experiments without clustering was set to $A_{0}$. We used $B=1$ for all experiments. The SA and PT parameters shown were used at each clustering step. For experiments without clustering, the number of iterations was multiplied by the number of clustering steps to ensure both methods were run for equal total update cycles. The inverse temperature ranges for both the SA and PT simulations were linearly separated between $0.0001 - 0.01$.

\begin{table}[htbp]
    \centering
    \caption{Hyperparameters used for the TSP simulations.}
    \label{tab:tsp-hparams}

    \setlength{\tabcolsep}{3pt}   
    \renewcommand{\arraystretch}{1.15}

    \begin{tabularx}{\textwidth}{l
        >{\centering\arraybackslash}p{0.8cm}   
        >{\centering\arraybackslash}p{0.7cm} >{\centering\arraybackslash}p{0.8cm} 
        >{\centering\arraybackslash}p{0.7cm} >{\centering\arraybackslash}p{0.8cm} 
        >{\centering\arraybackslash}p{0.7cm} >{\centering\arraybackslash}p{0.8cm} 
        >{\centering\arraybackslash}p{0.7cm} >{\centering\arraybackslash}p{0.8cm} 
        >{\centering\arraybackslash}p{1.1cm} >{\centering\arraybackslash}p{1.1cm} 
        >{\centering\arraybackslash}p{1.1cm} >{\centering\arraybackslash}p{0.9cm} >{\centering\arraybackslash}p{0.9cm}} 
        \toprule
        \multirow{2}{*}{\textbf{Instance}} &
        \multirow{2}{*}{\textbf{$A_0$}} &
        \multicolumn{8}{c}{\textbf{Clustering hyperparameters}} &
        \multicolumn{2}{c}{\textbf{SA}} &
        \multicolumn{3}{c}{\textbf{PT}} \\
        \cmidrule(lr){3-10} \cmidrule(lr){11-12} \cmidrule(lr){13-15}
        & &
        \textbf{$K_1$} & \textbf{$A_1$} &
        \textbf{$K_2$} & \textbf{$A_2$} &
        \textbf{$K_3$} & \textbf{$A_3$} &
        \textbf{$K_4$} & \textbf{$A_4$} &
        \textbf{Steps} & \textbf{Iters.} &
        \textbf{Iters.} & \textbf{Swap} & \textbf{Repls.} \\
        \midrule
        burma14   & 1000 & 4  & 1400 & -  & -    & -  & -    & -  & -  & 200 & 1000  & 10000 & 100 & 20 \\
        ulysses16 & 1500 & 8 & 2500 & 4 & 3000 & - & - & - & -  & 200 & 1000  & 10000  & 100 & 20 \\
        ulysses22 & 1500 & 16  & 2000 & 8  & 2500 & 4 & 3000 & - & - & 200 & 1000  & 10000  & 100 & 20 \\
        berlin52  & 1000 & 32  & 1000 & 16 & 1500 & 8 & 1500 & 4 & 2000 & 200 & 1000 & 10000 & 100 & 20 \\
        \bottomrule
    \end{tabularx}
\end{table}

\subsection{Spin-glass problem}
\label{subsec:app-sg-hyperparams}
Finally, for the spin-glass instances, the hyperparameters are shown in Table \ref{tab:parameters-SG-PT-SA}. We include the parameters, PT-baseline, which are the parameters used in the large PT run to estimate the baseline energies. 

\begin{table}[htbp]
    \centering
    \caption{The hyperparameters used for spin-glass simulations, including the baseline calculations.}

\begin{tabular}{ccccccccccccccc}
    \toprule
    \multirow{2}{*}{\textbf{Graph size}}
      & \multicolumn{1}{c}{\textbf{SA}}
      & \multicolumn{3}{c}{\textbf{PT}} 
      & \multicolumn{4}{c}{\textbf{PT-baselines}} \\
    \cmidrule(lr){2-2} \cmidrule(lr){3-5}
    \cmidrule(lr){6-10}
      & \textbf{Temp. range} 
      &\textbf{Temp. range}
      &\textbf{Swaps}
      & \textbf{Repls.}
      &\textbf{Temp. range}
      &\textbf{Swaps}
      & \textbf{Repls.}
      & \textbf{Iters.}
      & \textbf{Reps.}\\\\
    \midrule
    $N=100, 250, 500$  
      & $0.074- 0.74$   
      & $1.0-5.0$
      & $10$  
      & $5$
      & $1.0-5.0$  
      & $10$ 
      & $10$
      & $3000$ 
      & $20$\\
    $N=750, 1024$ 
      & $0.074- 0.74$     
      & $1.0-5.0$  
      & $10$  
      & $10$
      & $0.3-10$  
      & $10$
      & $20$ 
      & $3000$
      & $20$\\
    \bottomrule
\end{tabular}
\label{tab:parameters-SG-PT-SA}
\end{table}

\section{Strategies for implementing a VCPC in hardware}
\label{app:hardware_strategies}

An efficient implementation of a VCPC relies on a hybrid optoelectronic architecture, as detailed in Ref.~\cite{Ramy_2025}. Unlike conventional Ising machines that rely on physical coupling between components (such as capacitive coupling in CMOS or inductive coupling in superconducting loops), the VCPC decouples the stochastic bit generation from the network topology. The hardware consists of three primary subsystems: (1) a high-bandwidth photonic entropy source, (2) a high-speed digital logic unit (such as an FPGA or ASIC) for bias (update drive) calculation, and (3) memory banks for state and weight storage.
\paragraph{Entropy generation}
The stochasticity of the p-bits is derived from a photonic Quantum Random Number Generator (QRNG). The optical signal is detected and digitised to provide a continuous stream of random numbers, $rand_U[-1, +1]$, which are distributed to the update logic. The self-correcting nature and ultra-high speed of the entropy source allows for high-quality randomness without the need to engineer microscopic noise sources at every individual node.
\paragraph{Virtual connectivity and bias calculation}
The definition of the problem geometry is stored entirely in the digital memory. For a system of $N$ p-bits, the connectivity matrix $J$ (or tensor for higher-order problems) is stored in high-throughput memory (e.g., high bandwidth memory or on-chip SRAM). The calculation of the bias for a specific p-bit is performed by the digital logic unit.

To update a p-bit $i$, the logic unit retrieves the current states $\mathbf{m}$ of all connected neighbours and the corresponding coupling weights. The computing unit then performs a dot-product operation:
\begin{equation}
    I_i = \sum_{j} J_{ij} s_j + h_i.
\end{equation}
Crucially for the timing estimates in Sec.~\ref{subsubsec:SG-TTS}, this summation is implemented in hardware using a parallel adder tree architecture. In an adder tree, pairs of terms are summed in parallel at each stage, reducing the number of values to be summed by half at each level of the tree. Consequently, the depth of the logic circuit—and thus the number of clock cycles required to compute the sum—scales logarithmically with the number of connections, i.e., $O(\log_2 N)$ for a fully connected graph.

\paragraph{Parallel updates and clock frequency}
As discussed in Sec.~\ref{sec:overview}, graph colouring allows independent sets of p-bits to be updated simultaneously. The system operates at a clock frequency $f$ (e.g., 2.7 GHz as cited in Ref.~\cite{Ramy_2025}). Since the connectivity is virtual, changing the problem geometry (reconfigurability) or increasing the interaction order (higher-order problems) does not require physical rewiring; it requires only an update to the weights stored in memory and the routing logic in the adder tree.
This architecture directly informs the TTS derived in Sec.~\ref{subsubsec:SG-TTS} where the time per iteration is dominated by the latency of the adder tree depth.

\section{Derivation of the Hamiltonians and update drives}
\label{app:bias-eqns}
\subsection{Hitting set problem}
\label{app:HS}
\subsubsection{QUBO formulation}
\label{app:HS-Hamiltonian}
In the hitting set problem, we seek a subset of vertices $S \subseteq V$ of minimum size such that every hyperedge $r \in \mathcal{R}$ contains at least one vertex from $S$. We represent this with binary variables $s_v \in \{0, 1\}$, where $s_v = 1$ if vertex $v \in S$ and $s_v = 0$ otherwise. The constraint that every hyperedge $r$ must be covered is equivalent to requiring that it is not the case that all vertices in $r$ are outside of $S$. Mathematically, for each $r \in \mathcal{R}$, we require:
\begin{equation}
    \prod_{v \in r} (1 - s_v) = 0.
\end{equation}
To enforce this as an optimisation problem, we assign a large penalty $A$ to any violated constraint and minimise the total size of the set with weight $B$:
\begin{equation}\label{eqn:app-HS-Hamiltonian}
    E(\mathbf{s}) = A \sum_{r \in \mathcal{R}} \prod_{v \in r} (1 - s_{v}) + B \sum_{v \in V} s_v.
\end{equation}
Setting $A > B$ ensures that any valid vertex cover (where the first term is zero) will always have a lower energy than any invalid configuration.

\subsubsection{Update drive}
\label{app:HS-UpdateDrive}

We deduce from Eqn.~\ref{eqn:app-HS-Hamiltonian} that the update drive for the hitting set problem is:
\begin{equation}
\begin{split}
    I_k &= E(\mathbf{s}\vert s_k=0) - E(\mathbf{s}\vert s_k=1)\\
    &= A \left[ \sum_{\substack{r \in \mathcal{R} \\ k \notin r}} \prod_{v \in r} \left(1 - s_{v} \right) + \sum_{\substack{r \in \mathcal{R} \\ k \in r}} \prod_{\substack{v \in r \\ v \neq k}} \left(1 - s_{v} \right) \right] + B \sum_{\substack{v \in V(G)\\ v \neq k}} s_v \\& \quad - A \left[ \sum_{\substack{r \in \mathcal{R} \\ k \notin r}} \prod_{v \in r} \left(1 - s_{v} \right) \right] - B \sum_{\substack{v \in V(G)\\ v \neq k}} s_v - B \\
    & = A \sum_{\substack{r \in \mathcal{R} \vert \\ k \in r}} \prod_{\substack{v \in r \vert \\ v \neq k}} \left(1 - s_{v} \right) - B
\end{split}
\label{eqn:bias-derivation-HS}
\end{equation}
Note the dependence on the number of vertices in a hyperedge, or the dimension $k$, as well as the number of hyperedges, or $m$. 

\subsection{Travelling salesperson problem}
\subsubsection{QUBO formulation}
\label{app:TSP-Hamiltonian}
In order to formulate the TSP as a QUBO problem, we initially define a p-bit state matrix, $S$, which one-hot encodes our Hamiltonian cycle:
    \begin{equation}\label{eqn:S_ij}
    S = \begin{pmatrix}
        1 & 0 & 0 & \dots\\
        0 & 0 & 1 & \dots\\
        0 & 1 & 0 & \dots\\
        \vdots & \vdots & \vdots & \ddots
\end{pmatrix}
\end{equation}
The rows represent the city visited $\{A, B, C, ... \}$, and the columns denote the tour ordering. The example shown in Eqn.~\ref{eqn:S_ij} corresponds to the tour: $A\rightarrow C\rightarrow B$. Calculating the associated tour cost is straightforward using the distance matrix $D$:
\begin{equation}
E_{\text{cost}}
= \sum_{i=0}^{N-1}\sum_{j=0}^{N-1}\sum_{k=0}^{N-2} D_{ij}\, S_{ik}\, S_{j,k+1}
\;+\;
\sum_{i=0}^{N-1}\sum_{j=0}^{N-1} D_{ij}\, S_{i,N-1}\, S_{j,0}
\end{equation}
where the second term adds an implicit return to the starting city. We also need to introduce penalty terms to ensure $S$ converges to a valid tour. There are two types of invalid tours: (a) a tour which visits more or less than one city in a single visit and (b) a tour that visits the same city more or less than once. These constraints take the form:
\begin{equation}
    E_{\text{pen}} = \sum_{i=0}^{N-1}\bigg(\sum_{k=0}^{N-1}S_{ik}-1\bigg)^{2} + \sum_{k=0}^{N-1}\bigg(\sum_{i=0}^{N-1}S_{ik}-1\bigg)^{2}
\end{equation}
where the first and second terms correspond to constraints (a) and (b) respectively. Therefore, one can define the total energy function, $E$, for a TSP by linearly combining $E_{\text{pen}}$ and $E_{\text{cost}}$:
\begin{equation}\label{eqnapp:TSP-Hamiltonian}
\begin{aligned}
E(S)
&= A\Bigg[\sum_{i=0}^{N-1}\bigg(\sum_{k=0}^{N-1}S_{ik}-1\bigg)^{2}
   + \sum_{k=0}^{N-1}\bigg(\sum_{i=0}^{N-1}S_{ik}-1\bigg)^{2}\Bigg] \\
&\!+\, B\Bigg[\sum_{i=0}^{N-1}\sum_{j=0}^{N-1}\sum_{k=0}^{N-2}
   D_{ij}\, S_{ik}\, S_{j,k+1}
   + \sum_{i=0}^{N-1}\sum_{j=0}^{N-1} D_{ij}\, S_{i,N-1}\, S_{j,0}\Bigg]
\end{aligned}
\end{equation}
where $A$ and $B$ are hyperparameters. The ratio between $A$ and $B$ specifies the comparative importance of minimising the tour cost and ensuring the solver yields a valid tour. Conventionally, we set $B=1$ and $A$ is instance dependent, which is set subject to $A\geq\text{max}(D)$~\cite{Lucas_2014}. Due to the competitive nature of the terms comprising $E$, it is practically difficult to find the minimum tour using a choice of $A$ which consistently yields a valid tour. Moreover, it is well reported in Ref.~\cite{Dan_2020} that the probability of finding the optimal tour, using standard SA, approaches zero for problems involving $N>12$ cities. This necessitates additional procedures to assist the solver in finding the ground state.

\subsubsection{Update drive}
\label{app:TSP-UpdateDrive}
Using the TSP energy function from Eqn.~\ref{eqnapp:TSP-Hamiltonian} we can deduce the update drive:

\begin{equation}
\begin{split}
I_{ik} &= E(S_{ik}=0) - E(S_{ik}=1)\\
&= A\left[\left(\sum_{\substack{k'=0 \\ k'\neq k}}^{N-1} S_{ik'} - 1\right)^{2} 
        - \left(\sum_{\substack{k'=0 \\ k'\neq k}}^{N-1} S_{ik'}\right)^{2}\right] \\
&\quad+ A\left[\left(\sum_{\substack{i'=0 \\ i'\neq i}}^{N-1} S_{i'k} - 1\right)^{2} 
        - \left(\sum_{\substack{i'=0 \\ i'\neq i}}^{N-1} S_{i'k}\right)^{2}\right] \\
&\quad- B \Bigg[ \sum_{j=0}^{N-1} D_{ij}\, S_{j,k+1}
      + \sum_{i'=0}^{N-1} D_{i'i}\, S_{i',k-1} \Bigg]
\end{split}
\label{eqn:bias-derivation-TSP}
\end{equation}

Note that we can infer the three colouring rules by inspecting each term in $I_{ik}$. The first and second terms forbid p-bits in the same row and column to be in the same update group, respectively. While the third term couples $I_{ik}$ to all variables in $S_{i,k\pm1}$, which means p-bits in adjacent columns are directly dependent and cannot be included in the same update group. 

\subsection{Spin-glass problem}
\label{app:SG}
\subsubsection{Ising to QUBO mapping}
\label{app:SG-Mapping}
To map the Ising Hamiltonian $\mathcal{H} = -\sum_{i < j} J_{ij} \sigma_i \sigma_j - \sum_i h_i \sigma_i$ with $\sigma_i \in \{-1, +1\}$ to a binary QUBO problem $s_i \in \{0, 1\}$, we use the transformation $\sigma_i = 2s_i - 1$:

\begin{equation}
\begin{split}
    E(\mathbf{s}) &= -\sum_{i < j} J_{ij} (2s_i - 1)(2s_j - 1) - \sum_i h_i (2s_i - 1) \\
    &= -\sum_{i < j} J_{ij} (4s_i s_j - 2s_i - 2s_j + 1) - \sum_i h_i (2s_i - 1) \\
    &= \sum_{i < j} (-4J_{ij}) s_i s_j + \sum_i \left( 2 \sum_{j \neq i} J_{ij} - 2h_i \right) s_i + C.
\end{split}
\end{equation}

Defining $E(\mathbf{s}) = \sum_{i < j} Q_{ij} s_i s_j + \sum_i b_i s_i + C$, the parameters map as:
\begin{align}
    Q_{ij} &= -4J_{ij} \\
    b_i &= 2 \sum_{j \neq i} J_{ij} - 2h_i,
\end{align}
where $C$ is the constant energy offset. Minimising $E(\mathbf{s})$ is equivalent to finding the ground state of the original Ising system.

\subsubsection{Update drive}
\label{app:SG-UpdateDrive}
Lastly, we show the update drive for the spin-glass problem:

\begin{equation}
\begin{split}
    I_k &= E(\mathbf{s}\vert s_k=0) - E(\mathbf{s}\vert s_k=1) \\
    &= \left( \sum_{i<j, i,j \neq k} Q_{ij} s_i s_j + \sum_{i \neq k} b_i s_i \right) - \left( \sum_{i<j} Q_{ij} s_i s_j + \sum_i b_i s_i \right)_{s_k=1} \\
    &= - \left( \sum_{j > k} Q_{kj} s_j + \sum_{i < k} Q_{ik} s_i + b_k \right) \\
    &= - \left( \sum_{j \neq k} Q_{kj} s_j + b_k \right).
\end{split}
\label{eqn:bias-derivation-SG}
\end{equation}

Clearly, the number of terms which needs to be calculated is highly dependent on the density of $Q$.

\section{Average group numbers for the hitting set problem}
\label{app:heat-plots}

An important factor in considering which hypergraphs are efficient to run with the VCPC is the average group size when partitioning p-bits into independent groups. Recall that for a graph, independent nodes are nodes which do not share an edge; we generalise this for hypergraphs by defining independent nodes to be any group of nodes which, among them, share no hyperedges. 

The number of iterations $\mathcal{I}$ required
to reach a solution quality $q$ is reduced by a factor $N/|G|$, or the
number of p-bits divided by the number of groups. Equivalently,
this speed-up factor is given by the average group size, $\bar{G}$. 

\begin{figure}[h]
    \begin{subfigure}{0.5 \linewidth}
        \centering\includegraphics[width=0.9\linewidth]{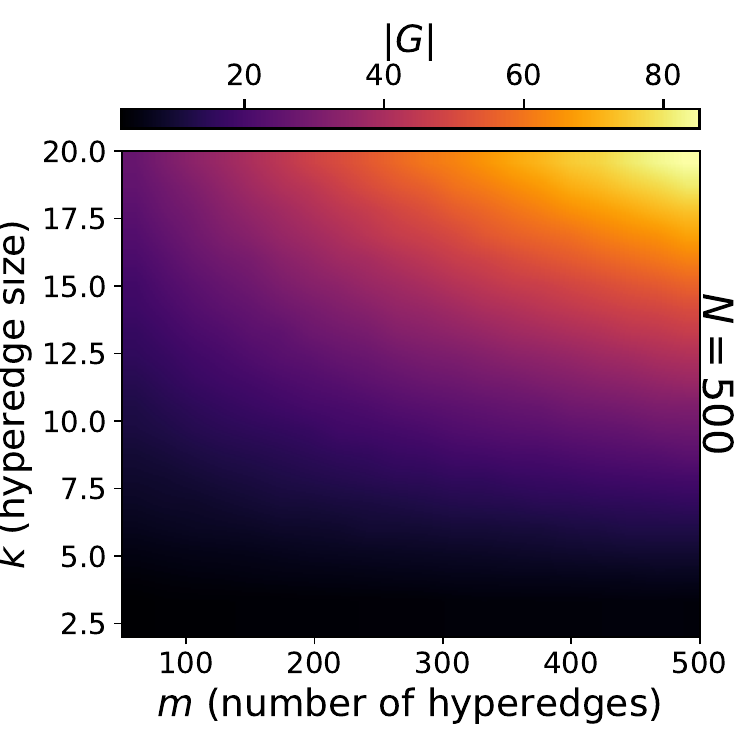}
        \caption{}
    \end{subfigure}
    \begin{subfigure}{0.5 \linewidth}
        \centering
        \includegraphics[width=0.9\linewidth]{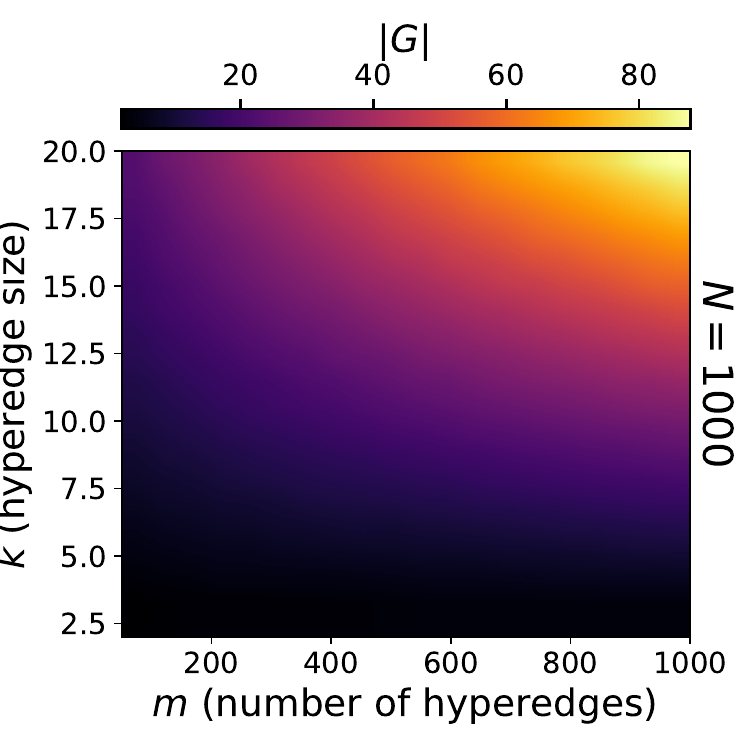}
        \caption{}
    \end{subfigure}
    \begin{subfigure}{0.5 \linewidth}
        \centering
        \includegraphics[width=0.9\linewidth]{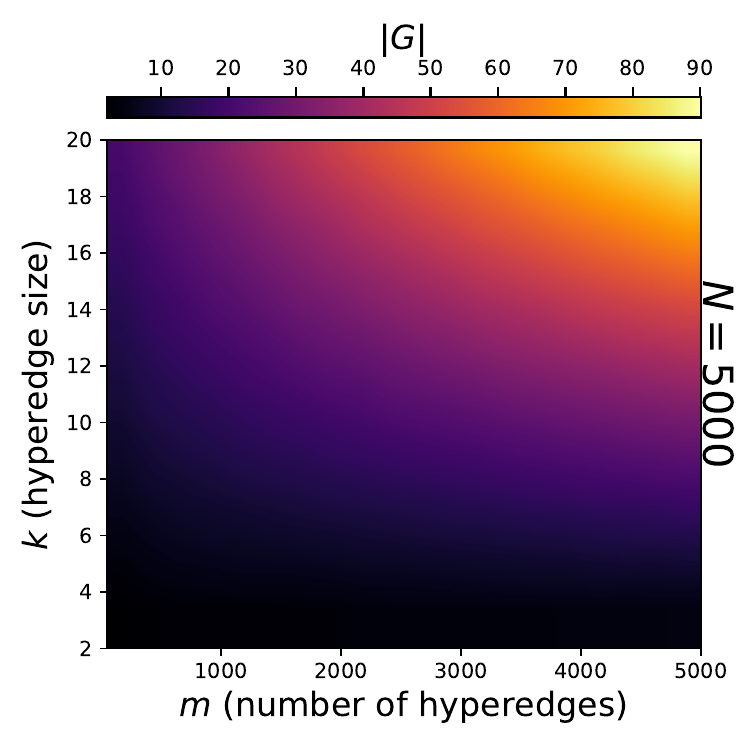}
        \caption{}
    \end{subfigure}
    \caption{Temperature plots showing the variation in the number of colour groups $|G|$ with the hypergraph dimension $k$ and the number of hyperedges $m$, for a fixed number of vertices $N$. The chosen values of $N$ were \textbf{(a)} $N=500$, \textbf{(b)} $N=1000$, and \textbf{(c)} $N=5000$.}
    \label{fig:heat-plot}
\end{figure}

In Fig.~\ref{fig:heat-plot} we plot temperature profiles of the number of groups $|G|$ for hypergraphs of size $N=500$, $1000$, and $5000$, for hypergraph dimensions in the range $[2, 20]$ and for varying numbers of hyperedges. Interestingly, they appear almost identical for the different hypergraph sizes; we can therefore safely assume that provided $k$ is not too large (size $6$ or less) for all values of $m$, the number of groups is less than $10$. As discussed in the main text, the factor of speed-up is then $\frac{N}{10}$, which for bigger hypergraphs, is clearly a larger speed-up due to group updates. This however must offset the increase in the time per iteration, incurred by the increased $m$ and $k$ for larger values of $N$.

\section{The effect of dimension on the conversion of the simulated annealing algorithm}
\label{app:HS-dimension}

\begin{figure}[h]
\centering
    \includegraphics[width = 0.7\textwidth]{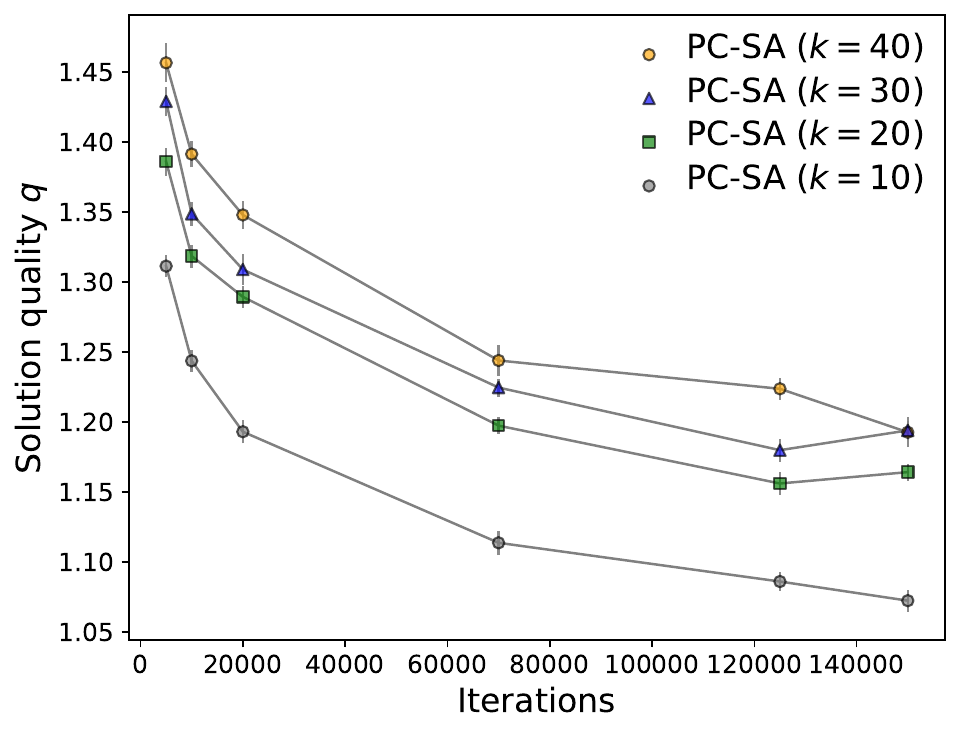}
    \caption{The average solution quality $q$ against iterations $\mathcal{I}$ obtained by the PC-SA algorithm for varying hypergraph dimensions. Dimensions $k=10$, $20$, $30$ and $40$ were tested. Other parameters, such as the number of vertices and edges of the hypergraph, were kept constant. Each point was averaged over the best obtained over $20$ repeats for $20$ hypergraphs.}
    \label{fig:k-comparison}
\end{figure}

We investigate how the hypergraph dimension $k$ of the hitting set problem affects the solution quality for the SA algorithm. Recall that increasing the dimension of the hypergraph corresponds to increasing the order of the energy function to be solved. To investigate this, in Fig.~\ref{fig:k-comparison} we plot the solution quality found by the PC-SA algorithm for hypergraphs of size $500$ vertices and $200$ edges, for varying values of $k$ and varying number of iterations. Note that group updates are not included here in the number of iterations. 

For each point we sample $20$ hypergraphs, and run $20$ repeats in parallel\footnote{We avoid calling these replicas, since for the case of PT replicas refers to the number of chains of different temperatures}. While for $k=10$ we see that the quality approaches $1.05$, for larger values of $k$ the solution quality is worse for an equal number of iterations. This is possibly due to the energy landscape becoming significantly more complex for higher-order problems, and therefore more difficult to sample.

\newpage
\section{Iterations against solution quality for the TSP problem}

\begin{figure}[h]
\centering
\begin{subfigure}{.49 \linewidth}
\centering
    \includegraphics[width=\textwidth]{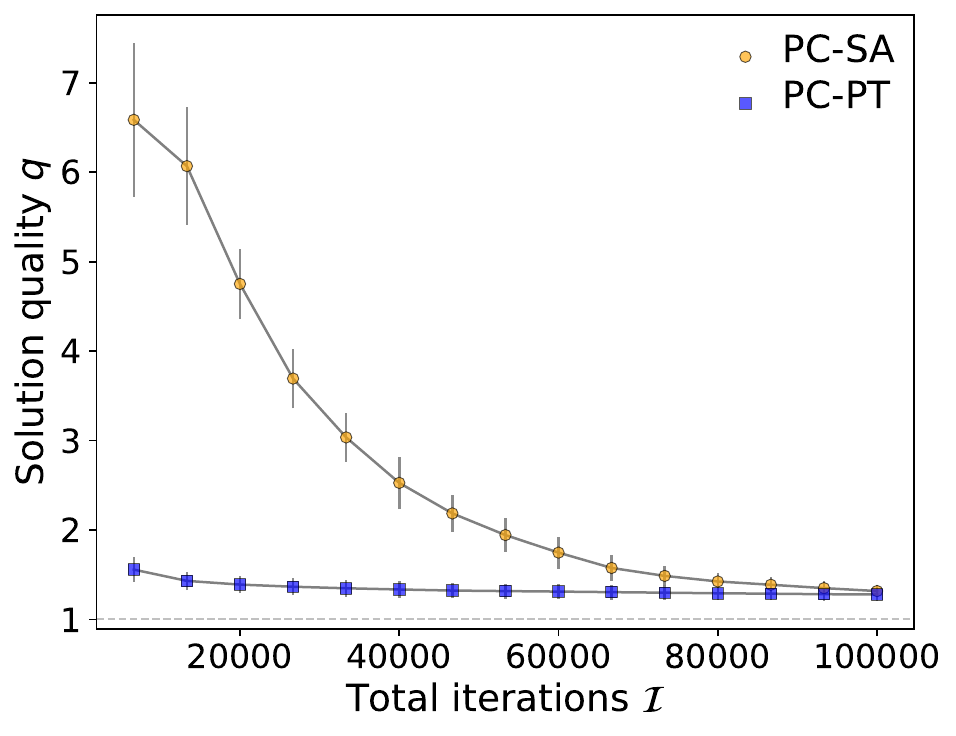}
    \caption{}
    \label{subfig:sol-quality-tspa}
\end{subfigure}
\begin{subfigure}{.49 \linewidth}
\centering
    \includegraphics[width=\textwidth]{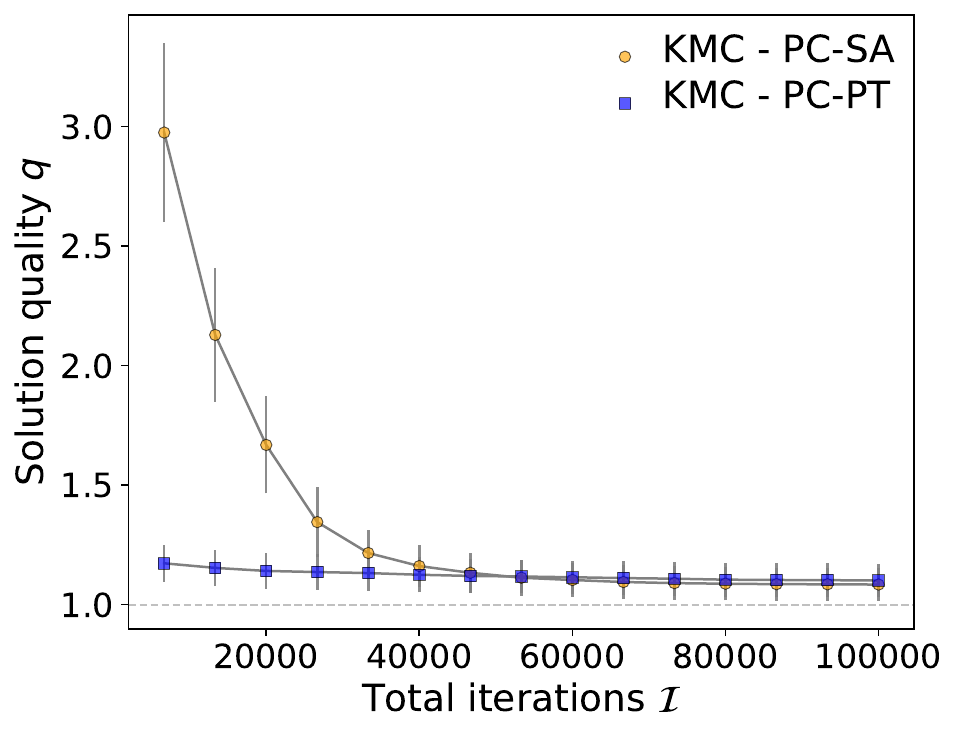}
    \caption{}
    \label{subfig:sol-quality-tspb}
\end{subfigure}
\caption{The best solution qualities at various numbers of total iterations $\mathcal{I}$ of the SA and PT algorithms without \textbf{(a)} and with \textbf{(b)} KMC. The error bars shows the range of solution qualities obtained for $100$ runs on the burma14 instance.}
    \label{fig:tsp-solution-quality}
\end{figure}

In Fig.~\ref{fig:tsp-solution-quality} we compare the number of iterations taken to reach various solution qualities for the burma14 instance with the four methods considered. We observe a similar trend in both Fig.~\ref{fig:tsp-solution-quality}a and Fig.~\ref{fig:tsp-solution-quality}b, where PC-PT is able to find lower solutions of a lower quality in several orders of magnitude fewer iterations than PC-SA. This is to be expected, as the lower temperature states are only explored by PC-SA during the latter iterations of the annealing process. The PC-PT algorithm however, is immediately exposed to these lower temperature states, and therefore can explore them constantly. Despite this initial improvement, on average the PC-SA reaches the best solution in less iterations than the PC-PT, both with and without KMC. This is most likely because the SA schedule evolves $\beta$ smoothly, allowing more consistent convergence toward low-energy states, whereas the rigid temperature exchanges in the $20$ replica PT can interrupt local relaxation.

\begin{figure}[h!]
\centering\includegraphics[width= 0.7 \textwidth]{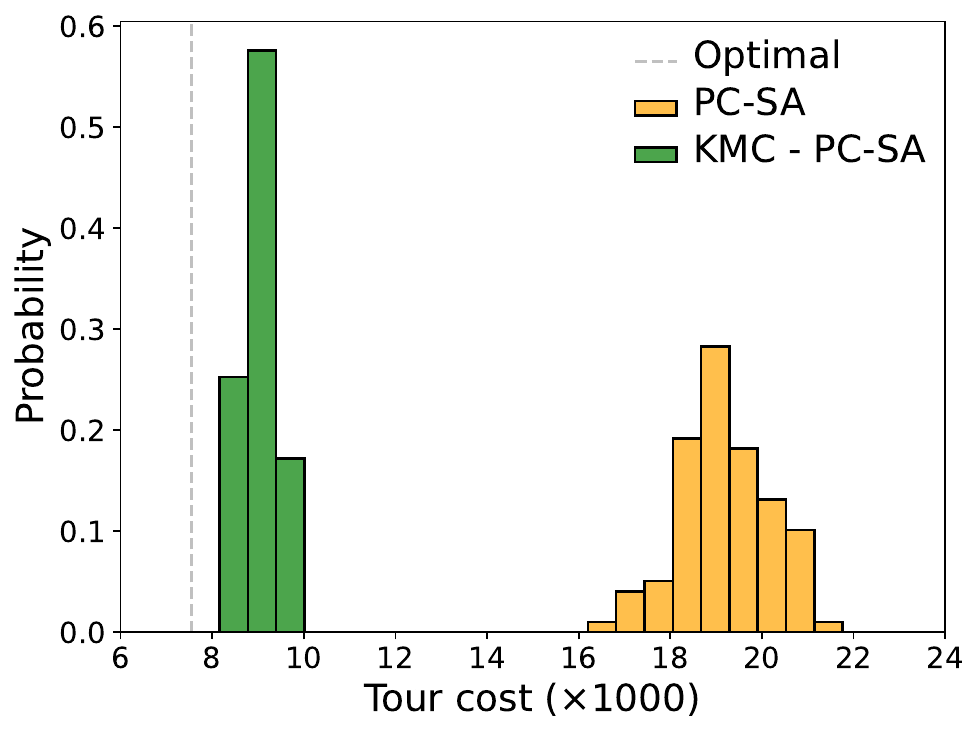}
\caption{Performances of standard PC-SA and KMC - PC-SA solvers on the berlin52 instance~\cite{Reinelt_1991}, where the optimal tour length is 7542.}\label{fig:berlin52-performance}
\end{figure}

We also show how the distribution of tours obtained changes with the addition of KMC in Fig.~\ref{fig:berlin52-performance}. The shift in the distribution shows clearly the advantage of employing KMC.

\clearpage